\documentclass[12pt]{article}
\usepackage{amssymb,amsmath}
\usepackage{graphicx}
\usepackage{epsfig}
\usepackage{epstopdf}
\usepackage{latexsym}
\usepackage{psfrag}
\usepackage{subfigure}
\usepackage{booktabs}
\usepackage{bbm}

\usepackage[toc]{appendix}

\usepackage{color}
\usepackage{datetime}
\usepackage[
      colorlinks=false,
      linkcolor=darkblue,  
      urlcolor=blue,    
      filecolor=blue,     
      citecolor=red,
linktocpage=true,
      pdfstartview=FitV,
      bookmarksopen=true    
      ]{hyperref}

\ifpdf
\fi

\numberwithin{equation}{section}

\renewcommand{\theequation}{\arabic{section}.\arabic{equation}}

\def\coeff#1#2{\relax{\textstyle {#1 \over #2}}\displaystyle}
\def\ds{\displaystyle}

\def\IR{\mathbb{R}}

\def\cN{{\cal N}}
\def\cO{{\cal O}}

\newcommand{\g}{\gamma}
\newcommand{\G}{\Gamma}

\newcommand{\ka}{\kappa}
\newcommand{\al}{\alpha}
\newcommand{\bb}{\beta}

\newcommand{\tht}{\theta}
\newcommand{\m}{\mu}

\newcommand{\bq}{\begin{equation}}
\newcommand{\eq}{\end{equation}}

\definecolor{cardinal}{rgb}{0.6,0,0}
\definecolor{darkgreen}{rgb}{0,0.5,0}
\definecolor{golden}{rgb}{0.92, 0.7, 0}
\definecolor{midnight}{rgb}{0, 0, 0.5}
\definecolor{darkblue}{rgb}{0.2, 0, 0.8}

\begin{document}  

\begin{titlepage}
 
\bigskip
\bigskip
\bigskip
\bigskip
\begin{center} 
{\Large \bf    Mind the Gap:  Supersymmetry Breaking in Scaling, Microstate Geometries}

\bigskip
\bigskip 

{\bf Orestis Vasilakis  and Nicholas P. Warner \\ }
\bigskip

Department of Physics and Astronomy \\
University of Southern California \\
Los Angeles, CA 90089, USA  \\
\bigskip
vasilaki@usc.edu,~warner@usc.edu  \\
\end{center}

\begin{abstract}

\noindent  We use a multi-species supertube solution to construct an example of a scaling microstate geometry for  non-BPS black rings in five dimensions. We  obtain the asymptotic charges of the microstate geometry and show how the solution is related to the corresponding non-BPS black ring.  The supersymmetry is broken in a very controlled manner using holonomy and this enables a close comparison with a scaling, BPS microstate geometry.     Requiring that there are no closed time-like curves near the supertubes places additional restrictions  on the moduli space of physical, non-BPS solutions when compared to their BPS analogs.    For large holonomy  the scaling non-BPS solution always has closed time-like curves while for smaller holonomy there  is a ``gap'' in the non-BPS moduli space relative to the BPS counterpart.  

\end{abstract}

\end{titlepage}


\tableofcontents

\section{Introduction}
 
One of the remarkable pieces of progress afforded by string theory, as a quantum theory of gravity,  is the possibility that one might be able to use it to describe the microstate structure of black holes.  At the perturbative level, in the limit of vanishing gravitational coupling this was demonstrated long ago for BPS black holes by Strominger and Vafa \cite{Strominger:1996sh}.   More recently, there has been progress at finite string coupling where the gravitational back-reaction is taken into account and the stringy object looks like a black hole (for reviews, see \cite{Mathur:2005zp,Bena:2007kg,Skenderis:2008qn,Balasubramanian:2008da,Chowdhury:2010ct}).   In the latter approach one strives to find and enumerate ``microstate geometries'' that have the same asymptotic behavior at infinity as a black hole and yet are smooth, horizonless geometries.   The idea is that such geometries might represent semi-classical microstates of the black hole, and this idea can be made quite precise if the black hole has a long $AdS$ throat  in which one can apply holographic methods.  The hope is that there will be sufficiently many microstate geometries to provide a representative sample of the actual quantum microstates of the black hole and perhaps yield a semi-classical description of some of the bulk thermodynamics, like the entropy.

Much of the progress on microstate geometries has also centered around BPS solutions because the supersymmetry of such backgrounds greatly simplifies the equations of motion.  However, in the last two years there has also been significant progress in constructing large families of non-BPS extremal solutions using ``Almost-BPS'' and ``floating-brane''   techniques \cite{Goldstein:2008fq, Bena:2009ev, Bena:2009en, Bena:2009fi}.   These constructions use the simplifications provided by supersymmetric systems and yet break the supersymmetry in a very controlled manner, typically using the holonomy of a background metric.   By using this approach one can readily reproduce, and then dramatically extend, the known extremal families of black-hole and black-ring solutions and one can find whole new families of solutions.

As yet, there are still rather few known non-BPS microstate geometries (some examples can be found in  \cite{Bena:2009qv, Bobev:2009kn, Bobev:2011kk}).  However,  we believe that the systematic construction of non-BPS microstate geometries is, at present,  primarily limited by the technical complexity of such solutions rather than by some strong physical limitation on their existence.   On the other hand, it is a very interesting and important physical question to investigate whether the breaking of supersymmetry does lead to limitations on the solutions, or on the moduli space of such solutions.   One of the purposes of this paper is to study precisely this phenomenon for a very simple microstate geometry corresponding to an ``Almost-BPS''  black ring.   

The simplicity of the supersymmetry-breaking mechanism in ``Almost-BPS'' solutions provides an ideal laboratory for studying this problem because one can start with supersymmetric configurations  and then turn on the supersymmetry breaking very slowly.  The microstate geometry that we will consider here will be a simple example of a ``bubbled black ring,''  in which the original, singular charge  sources have been replaced by smooth, cohomological fluxes supported on non-trivial cycles, or bubbles.  This results in a smooth, horizonless geometry that looks exactly like a black ring until one gets very close to the would-be horizon where the geometry caps off smoothly.  The bubbling process is, by itself,  $\frac{1}{8}$-BPS, preserving four supersymmetries locally, but in the Almost-BPS solution these supersymmetries are broken by the holonomy of the background. We find that, as the supersymmetry breaking holonomy gets stronger in the vicinity of the bubbled black ring,  the possible locations of the bubbles becomes progressively more limited and, in our simple example, one of the bubbles is required to become progressively smaller.   We also see that if the supersymmetry-breaking holonomy becomes too large then a physical solution ceases to exist\footnote{Specifically, there is no solution that  does not have closed time-like curves in the neighborhood of the bubbles.}.  We also contrast this with the corresponding supersymmetric, BPS bubbled black ring and see that no such restrictions occur.  

Thus we find that, compared to the BPS solution,  the Almost-BPS  object has a ``gap'' in its configuration space and the size of the gap increases with the strength of the supersymmetry breaking to a point where no physical solution exists.

While we are studying an extremely simple example in this paper, there are broader conclusions for microstate geometries.  The first and most evident is that supersymmetry breaking can restrict the configuration and moduli spaces.  Second, the restriction emerges from a competition between the supersymmetry breaking scale, which manifests itself,  in our example,  through the curvature scale and the scale of the fluxes in the bubbled geometries.  We find that the bubbles persist if the fluxes are large enough.  Conversely,  one might expect that if  one could ``dilute'' the supersymmetry breaking scale then bubbles with smaller fluxes would persist.  In our example, the supersymmetry breaking is generated by the holonomy of a Taub-NUT space and this supersymmetry breaking curvature can be diluted by passing to a multi-Taub-NUT.  The requirement that some bubbles have to be small might also be interpreted as requiring a solution to form ``locally supersymmetric'' clusters of bubbles.  The latter conclusions are rather speculative given the simplicity of our example but it does suggest some very interesting generalizations of our work here.

In Section 2 of this paper we outline the supergravity theory that we will study and give the equations that define a BPS and non-BPS systems of interest.  In Section \ref{Sols1} we specialize to a Taub-NUT background with three supertubes and we discuss why and how this represents a geometric microstate of a black hole in four dimensions.  We examine the regularity conditions and the requirements that there are no closed time-like curves (CTC's) and obtain the ``bubble equations'' or ``integrability conditions'' that constrain the supertube locations for both BPS and non-BPS solutions.  In Section \ref{ScalingSols} we examine the {\it scaling solutions}  in which the supertubes come very close together in the base geometry.  This limit corresponds to the opening of a deep black-hole, or black-ring, throat in the physical geometry and as a result, in this limit the microstate geometry looks more and more like a black object.  We give details of the asymptotic charges and structure of these geometries in Section  \ref{AsympStr}.  To clarify the differences between the BPS and non-BPS systems and  analyze the moduli space of solutions in more detail, we make some further simplifications  to the system of charges in the later parts of Section \ref{ScalingSols}.  We also solve the bubble equations in a flat space limit and exhibit two branches of solutions that will be part of our analysis of the more general solutions.

Sections \ref{LinBubb} and \ref{SolSpace} contain a careful analysis of the families of BPS and non-BPS solutions.  We first linearize the bubble equations around the coincidence limit of the points and show how the two branches of solutions persist in both the BPS and non-BPS solutions.  To find the additional restrictions  imposed by the non-BPS system it is easier to reverse the usual approach and, instead of fixing charges, we fix the positions of the supertubes and solve the bubble equations  for the charges.   We find that there are always physically sensible charges that solve the BPS bubble equations for any supertube location.  However, the non-BPS system has a non-trivial discriminant that restricts the locations of the supertubes:  We find a ``forbidden region,'' or ``gap,'' in the non-BPS moduli space.

Section \ref{Conclusions} contains our conclusions and discussion of the implications of our work.  Some technical aspects of the paper have been relegated to an appendix.

\section{The  family of non-BPS solutions}

As usual, it is simplest to characterize our solutions in terms of $\cN \! = \!  2$, five-dimensional ungauged supergravity coupled to two vector multiplets and we will use the conventions of \cite{Bena:2009fi}.  The bosonic sector of this theory  has three $U(1)$ gauge fields and two scalar fields and their action is given by:
\begin{eqnarray}
  S = \frac {1}{ 2 \kappa_{5}} \int\!\sqrt{-g}\,d^5x \Big( R  -\coeff{1}{2} Q_{IJ} F_{\mu \nu}^I   F^{J \mu \nu} - Q_{IJ} \partial_\mu X^I  \partial^\mu X^J -\coeff {1}{24} C_{IJK} F^I_{ \mu \nu} F^J_{\rho\sigma} A^K_{\lambda} \bar\epsilon^{\mu\nu\rho\sigma\lambda}\Big) \,,
  \label{5daction}
\end{eqnarray}
with $I, J =1,2,3$.  The scalars, $X^I$, satisfy a constraint and it is convenient to introduce three other scalar fields, $Z_I$, to
parametrize the $X^I$ via:
\begin{equation}
X^1 X^2 X^3  = 1\,, \qquad  X^1    =\bigg( \frac{Z_2 \, Z_3}{Z_1^2} \bigg)^{1/3} \,, \quad X^2    = \bigg( \frac{Z_1 \, Z_3}{Z_2^2} \bigg)^{1/3} \,,\quad X^3   =\bigg( \frac{Z_1 \, Z_2}{Z_3^2} \bigg)^{1/3}  \,.
\label{XZrelns}
\end{equation}
The matrix that defines the kinetic terms can be written as:
\begin{equation}
  Q_{IJ} ~=~    \frac{1}{2} \,{\rm diag}\,\big((X^1)^{-2} , (X^2)^{-2},(X^3)^{-2} \big) \,.
\label{scalarkinterm}
\end{equation}
The third, independent scalar is introduced via the warp factors in the metric Ansatz:
\begin{equation}
ds_5^2 ~=~ -Z^{-2} \,(dt + k)^2 ~+~ Z \, ds_4^2  \,,
\label{metAnsatz}
\end{equation}
with 
\begin{equation}
Z ~\equiv~ \big( Z_1 \, Z_2 \, Z_3  \big)^{1/3}   \,.
\label{Zdefn}
\end{equation}

To have families of ``floating $M2$ branes''  \cite{Bena:2009fi}, the metric functions and scalars are related to the electrostatic potentials by taking the gauge fields to be given by:
\begin{equation}
A^{(I)}   ~=~  - Z_I^{-1}\, (dt +k) + B^{(I)}  \,,
\label{AAnsatz}
\end{equation}
where $B^{(I)}$ is a one-form on the base, $ds_4^2$.

Introducing magnetic two-from field strengths associated with the $B^{(I)}$ 
\begin{equation}
\Theta^{(I)} ~\equiv~ d B^{(I)}\,.  \label{Thetadefn}
\end{equation}
one then finds that the equations of motion of supergravity are satisfied \cite{Bena:2009fi} if the base metric, $ds_4$,  is Ricci-flat and the following first-order linear system is satisfied:
\begin{eqnarray}
\Theta^{(I)} &=&  \varepsilon \star_4 \Theta^{(I)}\,,  \label{BPSequation1} \\
\widehat \nabla^2 Z_{I} &=&  \frac{1}{2}\, \varepsilon \, C_{IJK} \star_{4} ( \Theta^{(J)}\wedge \Theta^{(K)})\,,  \label{BPSequation2} \\
dk ~+~ \varepsilon \star_4 dk &=&  Z_{I}\, \Theta^{I} \,. \label{BPSequation3}
\end{eqnarray}
where $C_{IJK} \ = |\epsilon_{IJK} |$, $\widehat \nabla^2$ is the four-dimensional Laplacian and $ \star_4$ denotes duality on the four-dimensional base.   One can choose either duality by taking $\varepsilon = \pm 1$ and one still generates a solution to the equations of motion.

If the base metric is, in fact, hyper-K\"ahler with a curvature tensor whose duality matches that of $\Theta^{(I)}$ specified in (\ref{BPSequation1}), then the holonomy of the base metric respects the supersymmetries of the branes corresponding to the background metric and the solution is, in fact, BPS, preserving four supersymmmetries   \cite{Gauntlett:2004wh,Bena:2004de}.  The nice observation in \cite{Goldstein:2008fq, Bena:2009ev} was that one could still get a solution to the equations of motion while breaking supersymmetry by solving  (\ref{BPSequation1})--(\ref{BPSequation3}) in a hyper-K\"ahler base whose curvature had opposite duality to $\Theta^{(I)} $.  These ``Almost BPS'' solutions were then generalized to bases that are merely Ricci-flat  \cite{Bena:2009fi} and large families of new, non-BPS solutions were obtained \cite{Bena:2009ev, Bena:2009en, Bena:2009qv, Bena:2009fi, Bobev:2009kn}
  
The beauty of this solution-generating technique is not only that the solutions are defined by a linear system, and so superposition of solutions is trivial, but also that supersymmetry breaking occurs in a very controlled manner.  The brane configurations that underlie these solutions preserve four supersymmetries but these are broken by the holonomy of the background and so the scale of the supersymmetry breaking is set by the curvature scale of the metric on $ds_4$.  For flat base metrics, the four supersymmetries are fully restored and the solution is BPS.

\section{Some three-charge multi-supertube solutions in five dimensions}
\label{Sols1}
 
 \subsection{Supertubes as microstate geometries}
\label{STmicro}

We are going to consider a system of three different species of supertube in the simplest of hyper-K\"ahler base metrics, namely the Taub-NUT background.   
In our formulation, a ``type $I$'' supertube   corresponds to allowing an isolated, singular magnetic source for $\Theta^{(I)}$ with singular electric sources  in $Z_J$ and $Z_K$ ($I, J, K$ all distinct) at the same location.   Such a solution is, of course, singular in five dimensions.  However, one can uplift this to the six dimensions using the vector potential, $A^{(I)}$, as a Kaluza-Klein field and the resulting geometry is then completely regular $D1$-$D5$ supertube geometry in the $IIB$ duality frame \cite{Lunin:2001jy,Lunin:2002iz}.  The singular sources in five dimensions become  smooth magnetic flux on a three-dimensional bubble in six dimensions.

The obvious problem is that one can only perform this Kaluza-Klein uplift with one species of supertube and thus a solution with three different species of supertube will always have singularities in six dimensions.  Given that one can always resolve any single supertube singularity one might, quite reasonably, take the attitude that a multi-species supertube background should be considered to be a microstate geometry.  On the other hand, there is  better way to show that such a viewpoint is correct: One can use spectral flow \cite{Bena:2008wt, Bena:2009fi} to show that  a multi-species supertube solution can be used to generate a physically equivalent and truly non-singular microstate geometry.  

More specifically, once one has resolved one species of supertube by uplifting to six dimensions, one can perform a coordinate transformation in the six-dimensional solution and reduce back down to a {\it smooth} five-dimensional geometry in which the Taub-NUT space has been replaced by a more complicated base geometry.  For BPS solutions this simply modifies the potential that appears in the Gibbons-Hawking base   \cite{Bena:2008wt} but for non-BPS solutions the four-dimensional base is replaced by a generically more complicated electro-vac background \cite{Bena:2009fi}.   The important point is that, from the five-dimensional perspective, spectral flow represents a highly non-trivial transformation of the solution space in which singular supertube configurations are replaced by smooth fluxes on new two-dimensional cycles in the four dimensional base space.  Moreover, spectral flow can be done successively with each different species of supertube.  The only cost is that with each spectral flow, the base geometry and fluxes become more complicated, particularly for non-BPS solutions.  Indeed, the result of  a sequence of  three spectral flows of an Almost-BPS solution was recently obtained indirectly in \cite{Dall'Agata:2010dy}  by performing six T-dualities.
 
Thus, the important point is that a multi-species supertube solution, while singular in five dimensions, can always be transformed into a {\it physically equivalent}\footnote{It is equivalent because spectral flows are induced by coordinate transformations in six dimensions.} smooth, horizonless solution in five dimensions.   For  Almost-BPS systems,   the resulting geometry is typically very complicated and so it is much easier to work with the multi-species supertube solution, as we will here, because it encodes the physically equivalent, albeit rather more complicated, true microstate geometry.

 \subsection{Supertubes in Taub-NUT}
\label{TNsupertubes}
  
\subsubsection{The Metric}
\label{TNmet}

The Taub-NUT metric has the Gibbons-Hawking form:
\begin{equation}
ds^2_4 ~=~  V^{-1}(d\psi + A)^2 ~+~  V \, (dx^2 + dy^2 +   dz^2)     \,,
\label{GHmet}
\end{equation}
with 
\begin{equation}
 V ~=~  h ~+~ {q \over r}   \,,
\label{TNVdefn}
\end{equation}
where $r^2 \equiv \vec y \cdot \vec y$ with  $\vec y \equiv (x,y,z)$. This metric is hyper-K\"ahler if 
\begin{equation}
 \vec \nabla V ~=~ \pm  \vec \nabla \times \vec A   \,,
\label{VAreln}
\end{equation}
where  $\nabla$ denotes the flat derivative of $\IR^3$.  
The Riemann tensor is self-dual or anti-self-dual depending on the choice of sign (\ref{VAreln}). We will take the positive sign throughout and adopt a set of conventions so that this choice  corresponds to a self-dual curvature.   To that end, we introduce frames:
\begin{equation}
\hat e^1~=~ V^{-{1\over 2}}\, (d\psi ~+~ A) \,,
\qquad \hat e^{a+1} ~=~ V^{1\over 2}\, dy^a \,, \quad a=1,2,3 \,,
\label{GHframes}
\end{equation}
and define duality using these in canonical order.

The Riemann-squared invariant for this metric is:
\begin{equation}
R_{\mu \nu \rho \sigma}\,  R^{\mu \nu \rho \sigma}\, ~=~ \frac{24 \, h^2 \, q^2}{(q + h\,r)^6}   \,,
\label{Riemsq}
\end{equation}
which measures the strength of supersymmetry breaking in the non-BPS solutions.  

\subsubsection{Magnetic fields}
\label{MagSrc}

To define the magnetic fields for the background, it useful to introduce the self-dual and anti-self-dual  two-forms:
\begin{equation}
\Omega_\pm^{(a)} ~\equiv~ \hat e^1  \wedge \hat
e^{a+1} ~\pm~ \coeff{1}{2}\, \epsilon_{abc}\,\hat e^{b+1}  \wedge \hat e^{c+1} \,, \qquad a =1,2,3\,.\
\label{twoforms}
\end{equation}
The two-forms:
\begin{equation}
\Theta_\pm ~ \equiv~  -\sum_{a=1}^3 \,  (\partial_a P_\pm)  \,  \Omega_\pm^{(a)} \,,
\label{harmtwoform}
\end{equation}
are then harmonic if and only if
\begin{equation}
P_+ ~ =~ V^{-1} \, H    \quad {\rm or} \quad  P_- ~=~ H\,;  \qquad \qquad    \nabla^2 H ~=~0  \,,   
\label{harmP}
\end{equation}
where  $\nabla^2$ denotes the Laplacian on $\IR^3$.   

The vector potentials, $B_\pm$ with $\Theta_\pm = d B_\pm$, are given by:
\begin{equation}
B_\pm  ~=~ P_\pm  (d\psi + A) ~+~ \vec \xi_\pm \cdot d \vec y \,, \qquad \vec \nabla \times \vec \xi_+ = - \vec\nabla H  \,, \qquad \vec \nabla \times \vec \xi_-  =V \vec\nabla H  -  H \vec\nabla V \,.
\label{Bpmpots}
\end{equation}

For a magnetic dipole source located at $z=a_j$ on the $z$-axis, one has 
\begin{equation}
H ~=~ \frac{k_j}{r_j}  \,,
\label{Mdip1}
\end{equation}
where $k_j$ measures the dipole strength and
\begin{equation}
r_j~\equiv~  \sqrt{x^2 + y^2 + (z-a_j)^2}\,.
\label{ridefn}
\end{equation}
One can measure the local dipole strength by taking the integral of $\Theta_\pm $ over a small sphere, $S_\epsilon^2 \subset \IR^3$, around the singular point and one finds:
\begin{equation}
 \int_{S_\epsilon^2} ~  \Theta_+ ~=~   -2\,\pi\,  k_j \,, \qquad  \qquad \int_{S_\epsilon^2} ~  \Theta_- ~=~  + 2\,\pi\, \Big(h + \frac{q}{a_j} \Big) \, k_j  \,.
\label{Thetaint}
\end{equation}
For the two-sphere at infinity one has 
\begin{equation}
 \int_{S_\infty^2} ~  \Theta_+ ~=~   -2\,\pi\,  k_j \,, \qquad  \qquad \int_{S_ \infty ^2} ~  \Theta_- ~=~  + 2\,\pi\, h \, k_j  \,.
\label{Thetaintinfty}
\end{equation}
The difference between (\ref{Thetaint}) and (\ref{Thetaintinfty}) arises because $S_ \epsilon^2$ and $S_\infty^2$   are not homologous and $\Theta_- $ has a non-trivial flux through the non-compact $2$-cycle running defined by $(r,\psi)$ for $0 \le r < \infty$.

The local dipole strength for non-BPS supertubes leads to the natural definition of ``effective'' dipole  charges
\begin{equation}
\hat k_j ~\equiv~   \Big(h + \frac{q}{a_j} \Big) \, k_j \,.
\label{effectivek}
\end{equation}
It was found in  \cite{Bena:2009en} that this definition was also very natural because it was an essential step in bringing the expression for the horizon area,  $J_4$, of non-BPS black rings into its canonical form.

\subsubsection{Supertube solutions}
\label{STsols}

A ``type I'' supertube, $I=1,2,3$, has a singular magnetic source for $\Theta^{(I)}$ and singular electric sources for $Z_J$ and $Z_K$, where $I,J,K$ are all distinct.  We will study an axisymmetric supertube configuration with one supertube of each type on the $z$-axis.  We therefore source the magnetic fields,  $\Theta^{(J)}$, with harmonic functions: 
\begin{equation}
K^I  ~=~ \frac{k_I}{r_I}  \,,
\label{Kfns}
\end{equation}
with $r_I$ given by (\ref{ridefn}) and without loss of generality we will assume that 
\begin{equation}
a_1 ~>~ a_2  ~>~a_3 \,.
\label{ajorder}
\end{equation}
For BPS supertubes one can write down the complete solution using the results of 
 \cite{Bena:2005va, Berglund:2005vb, Bena:2007kg} and for non-BPS supertubes one can use the results in \cite{Bena:2009en}.  In particular, one has, for the BPS supertubes:
\begin{eqnarray}
Z_1 &=& 1 ~+~ \frac{Q_2^{(1)}}{4\,r_2} ~+~ \frac{Q_3^{(1)}}{4\,r_3} ~+~ \frac{k_2\,k_3 }{ (h + \frac{q}{r})\,r_2\, r_3} \,,  \label{BPSZ1} \\
Z_2 &=&  1  ~+~ \frac{Q_1^{(2)}}{4\,r_1} ~+~ \frac{Q_3^{(2)}}{4\,r_3} ~+~ \frac{k_1\,k_3 }{ (h + \frac{q}{r})\,r_1\, r_3} \,,  \label{BPSZ2} \\
Z_3 &=&  1   ~+~ \frac{Q_1^{(3)}}{4\,r_1} ~+~ \frac{Q_2^{(3)}}{4\,r_2} ~+~ \frac{k_1\,k_2 }{ (h + \frac{q}{r})\,r_1\, r_2} \,. \label{BPSZ3}
\end{eqnarray}
while for the non-BPS supertubes one has:
\begin{eqnarray}
Z_1 &=& 1 ~+~ \frac{Q_2^{(1)}}{4\,r_2} ~+~ \frac{Q_3^{(1)}}{4\,r_3} ~+~ \Big(h + \frac{q \, r}{a_2 \, a_3} \Big)\,  \frac{k_2\,k_3 }{ r_2\, r_3} \,,  \label{nonBPSZ1} \\
Z_2 &=&  1  ~+~ \frac{Q_1^{(2)}}{4\,r_1} ~+~ \frac{Q_3^{(2)}}{4\,r_3} ~+~ \Big(h + \frac{q \, r}{a_1 \, a_3} \Big)\, \frac{k_1\,k_3 }{ r_1\, r_3} \,,  \label{nonBPSZ2} \\
Z_3 &=&  1   ~+~ \frac{Q_1^{(3)}}{4\,r_1} ~+~ \frac{Q_2^{(3)}}{4\,r_2} ~+~ \Big(h + \frac{q \, r}{a_1 \, a_2} \Big)\, \frac{k_1\,k_2 }{ r_1\, r_2} \,. \label{nonBPSZ3}
\end{eqnarray}
The $Q^{(I)}_j$ define the local electric charge source of species $I$ at point $j$.  For the BPS solutions these also give the electric charges at infinity, but for the non-BPS solutions there is also a contribution from the dipole-dipole interaction term.

As usual one writes the Ansatz for the angular-momentum vector, $k$:
\begin{equation}
k ~=~ \mu\, ( d\psi + A   ) ~+~ \omega
\label{kansatz}
\end{equation}
and one can solve (\ref{BPSequation3})  for $\mu$ and $\omega$.  The expressions for these functions are completely explicit and details may be found in  \cite{Bena:2005va, Berglund:2005vb, Bena:2007kg} and   \cite{Bena:2009en}.  For BPS solutions, $\mu$ is given by 
\begin{equation}
\mu ~=~ \coeff {1}{6} \, V^{-2} \, C_{IJK}\,   K^I K^J K^K ~+~  \coeff{1}{2}  \,V^{-1}  \, K^I L_I ~+~  M\,, 
\label{muBPS}
\end{equation}
where $M$ is another harmonic function which we will take to be
\begin{equation}
M  ~=~ m_\infty ~+~ \frac{m_0}{r} ~+~   \sum_{j=1}^3  \, \frac{m_j}{r_j}  \,.
\label{Mdefn}
\end{equation}
Thus for the system we are studying,
\begin{eqnarray}
\label{BPSmu}
&&\mu  ~=~ \frac{k_1k_2k_3}{r_1 r_2 r_3 V^2} ~+~    \frac{1}{2V}\bigg(\frac{k_1}{r_1} \bigg(1+\frac{Q^{(1)}_{2}}{4r_2}+\frac{Q^{(1)}_{3}}{4r_3} \bigg) ~+~ \frac{k_2}{r_2} \bigg(1+\frac{Q^{(2)}_{1}}{4r_1}+\frac{Q^{(2)}_{3}}{4r_3} \bigg) \\
&& ~+~\frac{k_3}{r_3} \bigg(1+\frac{Q^{(3)}_{1}}{4r_1}+\frac{Q^{(3)}_{2}}{4r_2} \bigg) \bigg)  ~+~ M\,.
\end{eqnarray}
For non-BPS solutions the expression for $\mu$ is rather more complicated:
\begin{eqnarray}
 \label{munonBPS}
\mu&=&  \sum_I k_I\mu_I^{(1)}~+~h\sum_I\sum_{j\neq I}\frac{Q^{(I)}_{j}k_I}{4}\mu_{Ij}^{(3)}~+~q\sum_I\sum_{j\neq I}\frac{Q^{(I)}_{j}k_I}{4}\mu_{Ij}^{(5)} \\ 
&& \qquad ~+~ k_1k_2k_3\left( h^2\mu^{(6)}+q^2\mu^{(7)}+qh\mu^{(8)}\right)~+~\mu^{(9)} \,,
\end{eqnarray}
where, following \cite{Bena:2009en}, the  $\mu^{(j)}$ are defined by:
\begin{eqnarray}
 \label{munonBPSdefns}
&& \mu_I^{(1)}~=~ \frac{1}{2r_I} \,, \qquad \mu_{Ij}^{(3)}~=~\frac{1}{2Vr_Ir_j} \,, \qquad \mu_{Ij}^{(5)}~=~\frac{r^2+a_Ia_j-2a_Ircos\tht}{2Va_I(a_j-a_I)r r_I r_j}\,,  \\
&& \mu^{(6)}~=~\frac{1}{Vr_1r_2r_3} \,, \qquad \mu^{(7)}~=~\frac{rcos\tht}{Va_1a_2a_3r_1r_2r_3} , \\ 
&& \mu^{(8)}~=~\frac{r^2(a_1+a_2+a_3)+a_1a_2a_3}{2Vra_1a_2a_3r_1r_2r_3} \,, \qquad \mu^{(9)}~=~\frac{M}{V} \,.
\end{eqnarray}
%
%

%
%
The remaining details of the solution, including the expressions for $M$ and $\omega$ are given in Section \ref{AsympStr}, where we discuss the connection of the three supertube solution with the one of a black ring.

\subsection{Constituent charges}
\label{ConstChgs}

One of our primary purposes in this paper will be to compare ``the same''  BPS  and non-BPS  supertube configurations and the corresponding solution spaces.  There are, however, two natural notions of  being ``the same:''  one can either arrange to have the configurations made out of the same number and type of branes or one can arrange the configurations to have the same bulk charges measured at infinity.  We will adopt the former perspective primarily because it seems more physically in keeping with the idea that we are taking some otherwise supersymmetric collection of branes and using the holonomy of the background to break the supersymmetry and then studying the effects on the physics of the solution.  The charges and angular momenta measured at infinity will thus rather naturally depend upon the supersymmetry breaking process.  

On a more practical level, we want to understand and elucidate the effects of supersymmetry breaking on the possible geometric transitions to bubbled microstate geometries.  The system of equations and constraints is far simpler to understand in terms of local quantities whereas the asymptotic charges not only depend upon more complicated algebraic combinations of these local charges but also depend upon the geometric layout.  Since we are going to find limitations on the geometric layout as a result of supersymmetry breaking, the study of the effects of supersymmetry breaking as a function of asymptotic charges becomes a formidably entangled problem.  It is thus far  
easier to work with fixed constituent charges.

As we saw from  equation  (\ref{Thetaint}), the local dipole charges are determined by  $k_j$ for BPS supertubes and by $\hat k_j$ for non-BPS supertubes.   We also noted that if one tries to compare BPS and non-BPS black rings then it is the  ``effective charges,'' $\hat k_j$, of the non-BPS object that replace the dipole charges, $k_j$, of the BPS object  in canonical physical quantities like the horizon area    \cite{Bena:2009en}.    More directly, the strength of the divergence of the Maxwell field measures the local  {\it constituent} charge of the object in terms of the underlying branes and this is why these quantities naturally appear in formulae that determine horizon areas and entropies.   Thus the same supertube configuration is obtained by fixing the $\hat k_j$'s of the non-BPS configuration to the same values as the  $k_j$'s for the BPS configuration.  

It is also evident from (\ref{Thetaintinfty}) that fixing the constituent charges results in different charges measured at infinity but, as we remarked in Section  \ref{MagSrc}, this difference in charge is related to non-localizable fluxes through non-compact cycles.  It is also worth noting that the relationship between local and asymptotic charges is just as much  an issue for the electric charges because the BPS electric charges measured at infinity arising from (\ref{BPSZ1})--(\ref{BPSZ3})  depend solely upon the $Q_j^{(I)}$ whereas the non-BPS electric charges measured at infinity arising from  (\ref{nonBPSZ1})--(\ref{nonBPSZ3})  involve dipole-dipole interactions and depend upon the geometric details.  

Another advantage of fixing the constituent magnetic charges of a configuration is that the corresponding local electric charges are then easy to identify and fix.   For example,  for BPS supertubes, (\ref{BPSZ1}) shows that as $r_2 \to 0$ one has:
\begin{equation}
Z_1   ~\sim~   \frac{1}{4\,r_2}  \, \Big[\,  Q_2^{(1)} +  \frac{k_2\,k_3 }{ (h + \frac{q}{a_2}) \, |a_2 - a_3|} \,\Big] \,,
\label{BPSZ1div}
\end{equation}
whereas for non-BPS supertubes (\ref{nonBPSZ1})  yields:
\begin{equation}
Z_1   ~\sim~   \frac{1}{4\,r_2}  \, \Big[\,  Q_2^{(1)} + \Big(h + \frac{q }{a_3} \Big)\,  \frac{k_2\,k_3 }{ |a_2 - a_3|} \,\Big]   ~=~    \frac{1}{4\,r_2}  \, \Big[\,  Q_2^{(1)} + \frac{\hat k_2\,\hat k_3 }{(h + \frac{q }{a_2})\   |a_2 - a_3|}     \,\Big] \,.
\label{nonBPSZ1div}
\end{equation}
The electric charge thus has a pure source contribution, defined by the $Q_j^{(I)}$'s, and a part that comes from the magnetic dipole-dipole interactions.  More importantly, the local electric charges arising from the magnetic dipole-dipole interactions are identical between BPS and non-BPS solutions if one fixes $\hat k_j$ of the non-BPS solutions to the values of $k_j$ in the BPS solutions.   Thus the constituent charges of the BPS system are completely determined by $(Q_j^{(I)},k_j)$ and  by $(Q_j^{(I)},\hat k_j)$ for the non-BPS system and it is these sets of charges that are to be identified in order to get ``the same'' underlying local configuration.  In terms of string theory, this amounts to requiring the local brane constituents to be identical between BPS and non-BPS system. 

\subsection{Supertube regularity}

Supertubes are not regular in five dimensions but a ``type $I$'' supertube can be made regular in six dimensions, in the IIB frame by taking its Maxwell field and realizing it in terms of geometry as a Kaluza-Klein field. For the type $3$ supertube, the six-dimensional metric in IIB frame can be written as:
\begin{equation}
ds_6^2 ~=~ - \ds\frac{1}{Z_3\sqrt{Z_1Z_2}}\, (dt+k)^2  + \sqrt{Z_1Z_2}\,\,ds^2_4 + \ds\frac{Z_3}{\sqrt{Z_1Z_2}}\, (dz+A^{(3)})^2 \,. \label{sixmetric}
\end{equation}
where $A^{(3)}$ is the gauge potential defined in (\ref{AAnsatz}).   There are two things that need to be verified for supertube regularity:  One must first ensure that there are no divergent terms  along the $\psi$-fiber and then one must remove any CTC's associated with Dirac strings. Once one has done this the metric is completely regular at the supertube  \cite{Lunin:2001jy,Lunin:2002iz}.

\subsubsection{The regularity conditions}
\label{ParamFix}

Collecting all the $(d\psi +  A)^2$ terms in (\ref{sixmetric}):
\begin{equation}
(Z_1 Z_2)^{-\frac{1}{2}} \, V^{-2} \left[ Z_3\, (V P^3_\pm)^2 ~-~ 2 \mu \,
V^ 2 P^3_\pm ~+~ Z_1 Z_2 V \right] \, (d\psi +  A)^2 \,,
\label{fibercoeff}
\end{equation}
where $P^3_+  = V^{-1} K^3$ for BPS supertubes and $P^3_-  =  K^3$ for non-BPS supertubes.
For regularity as $r_3 \to 0$, one must have:
\begin{equation}
\lim_{r_3 \to 0}\,  r_3^2 \left[  \, Z_3\, (V P^3_\pm)^2 ~-~ 2 \mu\, V^2 P^3_\pm ~+~
Z_1 Z_2 V \, \right] ~=~ 0 \,. \label{regconda}
\end{equation}

To determine the condition for no Dirac strings it is simplest to look at the equation for $\omega$.  For BPS solutions one has:
\begin{equation}
\vec \nabla \times \vec \omega ~=~  ( V \vec \nabla \mu ~-~ \mu \vec \nabla V ) ~-~  V\, \sum_{I=1}^3 \,  Z_I \, \vec \nabla \bigg({K^I \over V}\bigg) \,, 
\label{BPSomega}
\end{equation}
while for non-BPS solutions one has:
\begin{equation}
\vec \nabla \times \vec \omega ~=~ - \vec \nabla (V \mu)   ~+~  V\, \sum_{I=1}^3 \,  Z_I \, \vec \nabla K^I  \,.
\label{nonBPSomega}
\end{equation}
To avoid Dirac strings at $r_3 =0$  one must have no terms that limit, as $r_3 \to 0$,  to a constant multiplet of $ \vec \nabla (\frac{1}{r_3})$ in the source on the right-hand side.  This means that one must have
\begin{equation}
\lim_{r_3 \to 0}\,  r_3 \left[ \, \mu ~-~Z_3\,  P^3_\pm\, \right] ~=~ 0  \,.  \label{regcondb}
\end{equation}

One can then use either  (\ref{regconda}),  (\ref{regcondb}) or some combination of them, to fix $m_3$ in (\ref{Mdefn}).  The other, independent condition, is then most easily expressed by 
eliminating  $\mu$ from (\ref{regconda}) using (\ref{regcondb}) to obtain:
\begin{equation}
\lim_{r_3 \to 0}\,  r_3^2 \left[ \, V Z_1 Z_2  ~-~ Z_3 \, (V P^3_\pm)^2 \,\right] ~=~ 0  \,.\label{regcondc}
\end{equation}
Regularity of the other supertubes in their IIB frames imposes conditions parallel to (\ref{regcondb}) and (\ref{regcondc}) as $r_1, r_2 \to 0$.

As was noted in \cite{Bena:2009en}, a rather technical calculation involving the explicit forms of $\mu$ shows that the $m_j$ are fixed to be:
\begin{equation}
m_1 ~=~  \frac{Q_1^{(2)} \, Q_1^{(3)}  }{32\,k_1}  \,, \qquad m_2 ~=~  \frac{Q_2^{(1)} \, Q_2^{(3)} }{32\,k_2}  \,, \qquad m_3 ~=~  \frac{Q_3^{(1)} \, Q_3^{(2)}}{32\,k_3}  \,,   \label{mjres}
\end{equation}
in both the BPS and non-BPS solutions.   Regularity and the absence  of Dirac strings at the origin imposes  conditions on  $\mu$ and thus fixes the parameters $m_0$ and  $m_\infty$  in the function $M$ defined in (\ref{Mdefn}) .  For the BPS solution, regularity requires that  $\mu$ be finite as $r \to 0$ and the absence of Dirac strings at the origin imposes the stronger requirement that  $\mu \to 0$ as $r \to 0$.  We therefore find that for the BPS solution we must impose:
\begin{equation}
  m_0~=~ 0 \,,  \qquad  m_\infty ~=~  -\sum_{i=1}^3 \frac{m_i}{a_i} \,.
  \label{BPSds1}
\end{equation}
The  result for the non-BPS solution is rather less edifying and its general form  may be found in  \cite{Bena:2009en}.  The details for the three supertube system will be given in Section \ref{AsympStr}.

\subsubsection{The bubble equations}
\label{BubbEqns}

Of central importance here are the other regularity conditions (\ref{regcondc}), and the similar conditions for $r_j \to 0$ in general,  because these produce the  bubble equations, or integrability conditions that constrain the locations of the supertubes in terms of the charges.  Define the symplectic inner products, $\Gamma_{ij}  = - \Gamma_{ji}$ and $\widehat \Gamma_{ij}  = - \widehat \Gamma_{ji}$ by:
\begin{equation}
\Gamma_{12}  ~=~  k_1\, Q_2^{(1)}  - k_2\, Q_1^{(2)}   \,, \qquad  \Gamma_{13}  ~=~  k_1\, Q_3^{(1)}  - k_3\, Q_1^{(3)}   \,, \qquad\Gamma_{23}  ~=~  k_2\, Q_3^{(2)}  - k_3\, Q_2^{(3)}   \,,  \label{Gammadefns}
\end{equation}
and
\begin{equation}
\widehat \Gamma_{12}  ~=~  \hat k_1\, Q_2^{(1)}  - \hat k_2\, Q_1^{(2)}   \,, \qquad  \widehat\Gamma_{13}  ~=~ \hat k_1\, Q_3^{(1)}  -\hat k_3\, Q_1^{(3)}   \,, \qquad \widehat\Gamma_{23}  ~=~  \hat k_2\, Q_3^{(2)}  - \hat k_3\, Q_2^{(3)}   \,,  \label{hatGammadefns}
\end{equation}
where the $\hat k_j$ are the effective dipole charges defined in (\ref{effectivek}).  Then the  bubble equations for the BPS supertubes may be written:
\begin{eqnarray}
\frac{\Gamma_{12}}{|a_1 - a_2|} ~+~ \frac{\Gamma_{13}}{|a_1 - a_3|} &=&  \frac{1}{4} \, \frac{Q_1^{(2)} \, Q_1^{(3)}  }{k_1}  \Big(h + \frac{q}{a_1} \Big) ~-~ 4 \, k_1 \,,  \label{BPSbubble1} \\
\frac{\Gamma_{21}}{|a_1 - a_2|} ~+~ \frac{\Gamma_{23}}{|a_2 - a_3|} &=&  \frac{1}{4} \, \frac{Q_2^{(1)} \, Q_2^{(3)}  }{k_2}  \Big(h + \frac{q}{a_2} \Big) ~-~ 4 \, k_2 \,,   \label{BPSbubble2} \\
\frac{\Gamma_{31}}{|a_1 - a_3|} ~+~ \frac{\Gamma_{32}}{|a_2 - a_3|} &=&  \frac{1}{4} \, \frac{Q_3^{(1)} \, Q_3^{(2)}  }{k_3}  \Big(h + \frac{q}{a_3} \Big) ~-~ 4 \, k_3 \,. \label{BPSbubble3}
\end{eqnarray}
while the bubble equations for the non-BPS supertubes are:
\begin{eqnarray}
\frac{\widehat\Gamma_{12}}{|a_1 - a_2|} ~+~ \frac{\widehat\Gamma_{13}}{|a_1 - a_3|} &=&  \frac{1}{4} \, \frac{Q_1^{(2)} \, Q_1^{(3)}  }{\hat k_1}  \Big(h + \frac{q}{a_1} \Big) ~-~ 4 \, \hat k_1 ~-~ \epsilon_{123} \, \widehat Y \,,  \label{nonBPSbubble1} \\
\frac{\widehat\Gamma_{21}}{|a_1 - a_2|} ~+~ \frac{\widehat\Gamma_{23}}{|a_2 - a_3|} &=&  \frac{1}{4} \, \frac{Q_2^{(1)} \, Q_2^{(3)}  }{\hat k_2}  \Big(h + \frac{q}{a_2} \Big) ~-~ 4 \, \hat k_2 ~-~ \epsilon_{213}   \, \widehat Y \,,   \label{nonBPSbubble2} \\
\frac{\widehat\Gamma_{31}}{|a_1 - a_3|} ~+~ \frac{\widehat\Gamma_{32}}{|a_2 - a_3|} &=&  \frac{1}{4} \, \frac{Q_3^{(1)} \, Q_3^{(2)}  }{\hat k_3}  \Big(h + \frac{q}{a_3} \Big) ~-~ 4 \, \hat k_3 ~-~ \epsilon_{312}   \, \widehat Y\,, \label{nonBPSbubble3}
\end{eqnarray}
where 
\begin{equation}
\widehat Y ~\equiv~    \frac{4\, h\,q \, k_1 k_2 k_3 }{a_1 a_2 a_3} ~=~    \frac{4\, h\,q \, \hat k_1 \hat k_2 \hat  k_3 }{(q+ h a_1) (q+ h a_2)  (q+ h a_3) } \,. \label{Ydefn}
\end{equation}
and
\begin{equation}
\epsilon_{ijk} ~\equiv~   \frac{(a_i - a_j)\,  (a_i - a_k)}{|a_i - a_j|\,  |a_i - a_k|}  \,. \label{epsdefn}
\end{equation}
Thus the BPS and non-BPS bubble equations are nearly identical, except for the additional term defined by $\widehat Y$,  if one replaces  the $k_j$ of the BPS configuration with the $\hat k_j$ of the non-BPS configuration.  This is because the bubble equations depend upon the local constituent charges of the underlying branes. 

It is also very interesting to note that $\widehat Y^2$ is proportional to the geometric mean of the curvature invariant, (\ref{Riemsq}), evaluated at the supertubes.  Thus $\widehat Y$ directly measures the strength of the supersymmetry breaking in the vicinity of the supertubes.  Our purpose now is to see how this affects the space of solutions to these equations.

\section{Scaling solutions}
\label{ScalingSols}

One chooses boundary conditions with $Z_I $ going to constants at infinity so that the space-time is asymptotically flat and with a Taub-NUT background, the non-compact space time is $\IR^{3,1}$, or four-dimensional, at large scales.    In order for the microstate geometry to look like a black hole (or black ring) at larger scales, all the multi-centered parts of the solution must cluster to look like a concentrated object and around this cluster there must be an ``intermediate region''  in which the warp factors (or electrostatic potential functions), $Z_I$,  behave as:
\begin{equation}
Z_I  ~\sim~   \frac{Q_I}{r_c} \,, \label{Zinter}
\end{equation}
where $r_c$ is the radial coordinate measured from the center of the cluster.  This intermediate region then defines the black-hole (or black-ring) throat and since the radial part of the metric behaves like $\frac{dr_c}{r_c}$ in this region, the distance diverges logarithmically as the cluster gets more and more tightly packed.   In the intermediate region, the physical metric approaches that of $AdS_3 \times S^2$ and the area of the black-hole-like throat is then determined by the $Q_I$ in  (\ref{Zinter}) and is finite if all of the $Q_I$ are non-zero. Thus an apparently singular coincidence limit in the base geometry is not singular in the full geometry.  On the contrary, it represents the physically most interesting limit in which a finite-sized black-hole throat opens up and all the microstate  details then cut off the throat and resolve the geometry at an arbitrarily depth set by the (small) size of the cluster. 

For regular, bubbled geometries in five dimensions, the $Z_I$ functions are finite everywhere and so there is a very important scaling limit that must be taken in order for the black-hole throat to open up in the proper manner  \cite{Bena:2006kb, Bena:2007qc}.  For the supertube backgrounds, the divergence of the $Z_I$ at the supertubes,  (\ref{BPSZ1})--(\ref{nonBPSZ3}), automatically guarantees the correct behavior,   (\ref{Zinter}),  of the $Z_I$ as one approaches a cluster.  It is interesting to recall that spectral flow does not modify the physics of a solution and yet with three spectral flows \cite{Bena:2008wt,Bena:2009fi} one can convert all three supertube species into pure geometric bubbles in five-dimensions in which the $Z_I$'s remain finite at the geometric centers.  Thus the simple beauty of using three supertubes is that one gets the microstate geometry of a black hole of finite horizon area simply by arranging the clustering of the supertubes. 

We therefore wish to study the scaling limit of our solutions and see when the two sets of bubble equations, (\ref{BPSbubble1})--(\ref{BPSbubble3})  or (\ref{nonBPSbubble1})--(\ref{nonBPSbubble3}), allow the supertubes to form an arbitrarily tight cluster.  

\subsection{Clustered supertubes}
\label{Clusters}

The first thing to note about the bubble equations is that if one adds them then the left-hand sides cancel and so the sum of the right-hand sides must be zero.  For a cluster, one has $a_j \to R$ for some fixed $R$, and so one finds the ``radius relation'' for the cluster, which is to be identified with the radius relation of the black ring that asymptotically our three supertube system looks like. This determines the location, $R$, of the cluster in terms of the charges and for BPS solutions it is a simple linear equation in $R$:
\begin{equation}
 \bigg[ \, \frac{Q_1^{(2)} \, Q_1^{(3)}  }{k_1} + \frac{Q_2^{(1)} \, Q_2^{(3)}  }{k_2} +\frac{Q_3^{(1)} \, Q_3^{(2)}  }{k_3}\,   \bigg]\,\Big(h + \frac{q}{R} \Big) ~=~16\, (k_1 + k_2 +k_3)\,, \label{BPSRreln}
\end{equation}
while for the non-BPS solution one has\footnote{Note that for any ordering of points, the sum of the three $\epsilon_{ijk}$'s defined in  (\ref{epsdefn}) is always $1$.} 
\begin{equation}
 \bigg[ \, \frac{Q_1^{(2)} \, Q_1^{(3)}  }{\hat k_1} + \frac{Q_2^{(1)} \, Q_2^{(3)}  }{\hat k_2} +\frac{Q_3^{(1)} \, Q_3^{(2)}  }{\hat k_3}\,   \bigg]\,\Big(h + \frac{q}{R} \Big) ~=~16\, (\hat k_1 + \hat k_2 + \hat k_3) ~+~ \frac{16 \, h\,q \, \hat k_1 \hat k_2 \hat  k_3 }{(q+ h R)^3}  \,, \label{nonBPSRreln}
\end{equation}
with 
\begin{equation}
\hat k_i  ~\approx~   \Big(h + \frac{q}{R} \Big) \, k_j \,.
\label{approxhatk}
\end{equation}
Whether one writes (\ref{nonBPSRreln}) in terms of $k_j$ or $\hat k_j$, the non-BPS radius relation is a cubic in $R$.

It is also obvious that the left-hand sides of the bubble equations for clusters are potentially divergent while the right-hand sides are finite.  This means that one must either have $\Gamma_{ij} \to 0$ or, more interestingly, one can have  $\Gamma_{ij}$ finite but with scaling arranged so that the divergences can on the left-hand sides of the bubble equations.  We will focus on arrangements of charges that allow the latter but do not necessarily exclude the former.  In particular, we will consider configurations for which
\begin{equation}
|a_1 -a_2|  ~\sim~    \lambda \,  \Gamma_{12} \,, \qquad  |a_2 -a_3|  ~\sim~    \lambda \,  \Gamma_{23} \,, \qquad |a_1 -a_3|  ~\sim~  -\lambda \,  \Gamma_{13} \,,
\label{scalingcond}
\end{equation}
for some {\it small} parameter, $\lambda$.  In particular, for $\lambda>0$, this means that $\Gamma_{12},  \Gamma_{23}   > 0$  and $\Gamma_{13} <  0$.
%
%
This means that for a scaling solution satisfying (\ref{scalingcond}) one must have:
\begin{equation}
 \Gamma_{12}   ~+~   \Gamma_{23} ~+~ \Gamma_{13}  ~=~ 0\,.
\label{sumGamma}
\end{equation}

There are other permutations that yield scaling solutions but our purpose here is not to make an exhaustive classification but to study the differences between BPS and non-BPS solutions.  We will thus make exactly the same assumptions for the $\widehat \Gamma_{ij}$ and for the ordering of non-BPS supertube configurations.

\subsection{A simplified system}
\label{SimpSystem}

From our discussion is Section \ref{ConstChgs}, to get the same supertube configurations we need to identify $(Q_j^{(I)},k_j)$ for the BPS system with $(Q_j^{(I)},\hat k_j)$ for the non-BPS system.   To simplify things still further, we will take all the dipole charges to be exactly the same.    This means we take $k_j = d, j=1,2,3$ for BPS solutions and  $\hat k_j = d, j=1,2,3$ for non-BPS solutions and treat $d$ as a fixed dipole field strength in both instances. With this choice (\ref{sumGamma}) becomes 
\begin{equation}
Q_2^{(1)}   ~-~  Q_1^{(2)} ~+~ Q_3^{(1)}   ~-~  Q_1^{(3)} ~+~  Q_3^{(2)}   ~-~  Q_2^{(3)} ~=~ 0\,,
\label{sumchgs}
\end{equation}
for both the BPS and non-BPS systems.

While  (\ref{sumchgs}) means that there are five possible independent charges, we will keep things very simple by passing to the three-parameter subspace defined by:
\begin{equation}
Q_2^{(1)}  ~=~  Q_2^{(3)}  ~=~ \alpha\,, \qquad   Q_3^{(1)}   ~=~  Q_1^{(2)} ~=~ \beta\,, \qquad     Q_1^{(3)}   ~=~ Q_3^{(2)} ~=~   \gamma   \,.
\label{Chgchoice}
\end{equation}
With this choice one has, for the BPS system, as $r \to \infty$:
\begin{equation}
Z_1~\sim~ \frac{ \alpha +  \beta}{4\, r}  \,, \qquad     Z_2~\sim~ \frac{ \beta+  \gamma}{4\, r}  \,, \qquad  Z_3~\sim~ \frac{ \alpha +  \gamma}{4\, r}    \,,
\label{Zasympinfty}
\end{equation}
which means that $\alpha, \beta$ and $\gamma$ can be used to parametrize three independent electric charges at infinity.  The corresponding asymptotics are a little more complicated for the non-BPS system but the conclusion is still the same.

With these choices, the bubble equations reduce to
\begin{eqnarray}
\frac{ \alpha- \beta}{a_1 - a_2} ~-~ \frac{\gamma - \beta}{a_1 - a_3} &=&  \frac{1}{4} \, \frac{\beta \, \gamma }{d^2}  \Big(h + \frac{q}{a_1} \Big) ~-~ 4   ~-~   Y \,,  \label{simpbubble1} \\
- \frac{ \alpha- \beta}{a_1 - a_2} ~+~ \frac{\gamma - \alpha}{a_2 - a_3} &=&  \frac{1}{4} \, \frac{\alpha^2 }{d^2}  \Big(h + \frac{q}{a_2} \Big) ~-~ 4  ~+~ Y \,,   \label{simpbubble2} \\
\frac{\gamma - \beta}{a_1 - a_3} ~-~ \frac{\gamma - \alpha}{a_2 - a_3} &=&  \frac{1}{4} \, \frac{\beta \, \gamma}{d^2}  \Big(h + \frac{q}{a_3} \Big) ~-~ 4  ~-~  \,  Y\,, \label{simpbubble3}
\end{eqnarray}
where $Y=0$ for the BPS system and 
\begin{equation}
Y~=~ \widetilde Y ~\equiv~     \frac{4\, h\,q \, d^2 }{(q+ h a_1) (q+ h a_2)  (q+ h a_3) }  = \, \frac{\widehat Y}{d} \,,\label{Ytildedefn}
\end{equation}
for the non-BPS system.  Note that for $\lambda >0$ the scaling conditions (\ref{scalingcond}) are equivalent to $\gamma > \alpha >  \beta$.
%
%
As we noted earlier, the only difference between the BPS bubble equations and the non-BPS bubble equations is a source term related to the background curvature that is doing the supersymmetry breaking.

It is also useful to rewrite these equations by multiplying by the obvious common denominators.  One the obtains
\begin{eqnarray}
-(\gamma - \alpha) \, a_1  +   (\gamma - \beta) \, a_2 -  (\alpha- \beta) \, a_3  &=& (a_1 -a_2) (a_1 -a_3) \bigg(   \frac{1}{4} \, \frac{\beta \, \gamma }{d^2}  \Big(h + \frac{q}{a_1} \Big) - 4  - Y \bigg) \nonumber\\
&=& - (a_1 -a_2) (a_2 -a_3) \bigg(   \frac{1}{4} \, \frac{\alpha^2 }{d^2}  \Big(h + \frac{q}{a_2} \Big) -4  + Y \bigg) \nonumber   \\
&=& (a_1 -a_3) (a_2 -a_3) \bigg(   \frac{1}{4} \, \frac{\beta \, \gamma }{d^2}  \Big(h + \frac{q}{a_3} \Big) - 4 -Y \bigg) \,.\nonumber \\ \label{simpbubble4}
\end{eqnarray}
%
 
\subsection{The solution in flat, cylindrical geometry}
\label{Cylinder}

To understand the branches of the solution, it is very instructive to start by solving the equations with $q=0$ and subsequently study what happens as $q$ is reintroduced.  Setting $q=0$ reduces the geometry of the base space to a flat  cylinder, $\IR^3 \times S^1$, where the radius of the circle is set by the value of $h$.   The BPS and non-BPS bubble equations become identical and trivially solvable.  

First one should note that because the cylinder is translationally invariant, it is only the differences $(a_i -a_j)$ that are going to have physical meaning.  Taking the sum of the bubble equations yields
\begin{equation}
h ~=~ \frac{48\, d^2}{\alpha^2 +2\, \beta\, \gamma}  \,. \label{trivh}
\end{equation}
The radii  of the supertubes are determined by their charges and this identity forces the supertube radii to match that of the $S^1$ of the cylinder. 

The system of equations (\ref{simpbubble4})  has three branches of solution.   Taking the difference between (\ref{simpbubble1}) and (\ref{simpbubble3}) one easily finds that  there is a set of solutions that has $a_2$ mid-way between $a_1$ and $a_3$:
\begin{equation}
 \quad (a_1 -a_2) ~=~ (a_2 -a_3) ~=~  \frac{  (\alpha^2 + 2\, \beta\, \gamma)}{8\,(\alpha^2 - \beta\, \gamma)} \,(2\, \alpha -  \beta - \gamma)  \,, \label{arithbrnch1}
\end{equation}
We call this the {\it symmetric branch}.

The other two branches allow the $a_i$ to be placed at any position but the charges are constrained accordingly.
That is, we can solve  (\ref{simpbubble4}) for $\alpha, \beta$ and $\gamma$ in terms of the $a_i$ and we find either
\begin{equation}
 \label{quarticarithmetic}
\alpha~=~ \beta~=~ \gamma~=~   \pm\frac{4d}{\sqrt{h}} \,,
\end{equation}
or 
\begin{equation}
 \label{quarticgeometric}
\alpha=\pm\frac{4d}{\sqrt{h}}\,, \qquad  \beta=\pm\frac{4d|a_1-a_2|}{\sqrt{h}|a_2-a_3|}\,, \qquad \gamma =\pm\frac{4d|a_2-a_3|}{\sqrt{h}|a_1-a_2|} \,.
\end{equation}
We refer to the latter as the {\it geometric branch} because the ratios of charges are related to ratios of separations and because the charge $\alpha$ is the geometric mean of $\beta$ and $\gamma$:
\begin{equation}
 \alpha^2 - \beta\, \gamma ~=~ 0\,.   \label{geombrnch1}
\end{equation}

As we will see, the solution given by (\ref{quarticarithmetic}) extends, for $q\ne 0$, to a scaling solution with $\alpha \to \frac{1}{2} (\beta + \gamma)$ in the scaling  limit,  $a_i -a_j \to 0$.  We will therefore refer to such solutions as the  {\it arithmetic branch}.  The three branches of solution obviously meet when all the points and charges coincide.

One should also recall that we required:
\begin{equation}
\gamma ~>~  \alpha ~>~ \beta  \,,   \qquad  a_1 ~>~  a_2  ~>~ a_3  \,,  \label{reqinequal}
\end{equation}
Combining this with (\ref{quarticgeometric}) one finds the additional condition 
\begin{equation}
 \label{geomadditional}
a_2-a_3 ~>~ a_1-a_2
\end{equation}
on the geometric branch.
To arrange the supertube geometry so that $a_2-a_3<a_1-a_2$ we just have to flip the signs in (\ref{scalingcond}), which also flips the order of the charges $\gamma ~<~  \alpha ~<~ \beta$. This transformation leaves  (\ref{sumGamma}) unaffected.

It is evident that from (\ref{quarticarithmetic}) and  (\ref{quarticgeometric})  that, for $q=0$, the fluxes, $\Gamma_{ij}$, vanish on the arithmetic branch and remain finite on the geometric branch.   More generally, for the scaling limit with $q \ne 0$,  we will see in Section \ref{NumEx}  that one typically has  $\Gamma_{ij} \to 0$ on the arithmetic branch while $\Gamma_{ij}$ remains finite on the geometric branch.  In addition, one can only have $\alpha=\beta =\gamma$   if $q=0$ or $q \to \infty$.

\section{Linearizing the bubble equations}
\label{LinBubb}
 
To define the linearization of the bubble equations, we first define the parameter, $R$,  that sets the location at which all the supertubes would coinicide.  This is given by (\ref{BPSRreln}) or  (\ref{nonBPSRreln}) for the BPS and non-BPS systems respectively.  These collapse to:
\begin{equation}
  \frac{1}{4} \,\Big( \, \frac{ \alpha^2 + 2\,\beta \, \gamma }{d^2}\,   \Big) \,\Big(h + \frac{q}{R} \Big) ~=~12  ~+~ Y_0  \,, \label{simpRreln}
\end{equation}
where $Y_0 =0$ for the BPS system while for the non-BPS system one has:
\begin{equation}
Y_0~=~      \frac{4\, h\,q \, d^2 }{(q+ h R)^3 }  \,. \label{Y0defn}
\end{equation}

Having thus defined $R$,  we now consider a solution that is very close to this limit point and introduce a small parameter, $\lambda$, that is {\it defined} by setting $a_2 = (1 + \lambda) R$.  One also assumes that $(a_i - a_j) \sim \cO(\lambda)$.  A quick examination of the bubble equations shows that at first order one must have $(a_1 - a_2)  =  \lambda \mu (\alpha -\beta)$  and $(a_2 - a_3)  =  \lambda \mu (\gamma - \alpha)$ for some parameter, $\mu$.  Thus we are led to the expansion:
\begin{eqnarray}
 a_1    &=&   (1 + \lambda) \, R  ~+~  \lambda\, \mu\,  (\alpha -\beta) ~+~ \lambda^2\, \mu^2\,  x_1 \,,   \label{expansion1}\\
 a_2     &=&  (1 + \lambda) \, R   \label{expansion2}\,,   \\
 a_3    &=&   (1 + \lambda) \, R  ~-~  \lambda\, \mu\,  (\gamma - \alpha) ~+~ \lambda^2\, \mu^2\,  x_3\,,   \label{expansion3}
\end{eqnarray}
for some parameters, $\lambda, \mu, x_1$ and $x_3$.  One then finds that (\ref{simpbubble4}) is trivially satisfied at zeroeth and first order in $\lambda$ and the first, non-trivial set of equations emerges at second order:
\begin{eqnarray}
-(\gamma - \alpha) \, x_1    -  (\alpha- \beta) \, x_3  &=& (\alpha- \beta) (\gamma - \beta) \bigg(   \frac{1}{4} \, \frac{\beta \, \gamma }{d^2}  \Big(h + \frac{q}{R} \Big) - 4  - Y_0\bigg) \nonumber\\
&=& - (\alpha- \beta) (\gamma - \alpha) \bigg(   \frac{1}{4} \, \frac{\alpha^2 }{d^2}  \Big(h + \frac{q}{R} \Big) -4  + Y_0 \bigg) \nonumber   \\
&=& (\gamma - \alpha) (\gamma - \beta) \bigg(   \frac{1}{4} \, \frac{\beta \, \gamma }{d^2}  \Big(h + \frac{q}{R} \Big) - 4 -Y_0 \bigg) \,.\label{linear1}
\end{eqnarray}
The equality of the three right-hand-sides of these equations gives two conditions, one of them is the defining ``radius relation'' for $R$, given by (\ref{simpRreln}).  For the BPS system, the second condition reduces to:
\begin{equation}
   (\alpha- \beta) \,  (2 \alpha- \beta- \gamma)\,(\alpha^2 - \beta \, \gamma )   ~=~ 0 \,, \label{BPSreln1}
\end{equation}
while for the non-BPS system it becomes:
\begin{equation}
   (\alpha- \beta) \,  (2 \alpha- \beta- \gamma)\, \big((\alpha^2 + \beta \, \gamma ) \, (q + h\,R)  - 32\, R\, d^2 \big) ~=~ 0 \,. \label{nonBPSreln1}
\end{equation}

Both BPS and non-BPS systems admit the arithmetic branch, which naively involves setting $\alpha = \frac{1}{2} (\beta + \gamma)$, but remembering that we are making a series expansion, the arithmetic branch is defined more precisely by taking:
\begin{equation}
\alpha ~=~  \frac{1}{2}\, (\beta + \gamma) ~+~ \cO(\lambda) \,. \label{arithbrnch2}
\end{equation}
The geometric branch involves taking 
\begin{equation}
\alpha^2 ~=~ \beta \, \gamma  ~+~ \cO(\lambda) \,, \label{geombrnch2a}
\end{equation}
for the BPS system and 
\begin{equation}
(\alpha^2 + \beta \, \gamma ) \, (q + h\,R)  - 32\, R\, d^2  ~+~ \cO(\lambda) ~=~ 0 \,. \label{geombrnch2b}
\end{equation}
for the non-BPS system.

For both systems, the right-hand side of (\ref{linear1}) vanishes on the geometric branch and one has:
\begin{equation}
   \frac{x_1}{x_3} ~=~  - \frac{(\alpha- \beta)}{(\gamma - \alpha)}   \,, \label{geombrnch3}
\end{equation}
One can then use the fact that, to leading order, the right-hand side of (\ref{linear1}) vanishes to rewrite the condition (\ref{geombrnch3}) as
\begin{equation}
 \label{OreGeometric}
\alpha^2~-~ \beta\,\gamma ~=~ -\frac{32hqd^4}{(q + h\,R)^3}~+~ \cO(\lambda)
\end{equation}
which shows, more explicitly, how the BPS and non-BPS geometric branches differ.

One can expand to higher orders and obtain expressions for $\lambda$ and for how the $\cO(\lambda)$ terms in (\ref{arithbrnch2}), (\ref{geombrnch2a}) or  (\ref{geombrnch2b}) relate to the values of $x_1, x_3$ and higher order corrections.  We will not pursue this here but instead we will move on to discuss the restrictions placed upon the non-BPS space of solutions.  To that end, we note that (\ref{geombrnch2b}) is linear in $R$ and is trivially solvable to yield:
\begin{equation}
R ~=~ \frac{q\, (\alpha^2 + \beta \, \gamma)}{32\,d^2- h\, (\alpha^2 + \beta \, \gamma)}  +~ \cO(\lambda)   \,. \label{rreln2}
\end{equation}
One can now substitute this in the defining relation, (\ref{simpRreln}), for  $R$ and the result is a complicated polynomial relationship between $\alpha, \beta,\gamma$ and $d$ that is required for the non-BPS system.  This replaces the simple constraint (\ref{geombrnch2a}) on the charges that arises in the BPS system.  The non-trivial constraint on the charges for the non-BPS system is a quartic in $\alpha^2$ and $\beta \gamma$ and reduces to:
\begin{equation}
(\alpha^2 + \beta \, \gamma)\, \big(h\, (\alpha^2 + \beta \, \gamma) - 32\, d^2 \big)^3~-~32768\, q^2 \, d^4  \, (\alpha^2 -  \beta \, \gamma)  ~=~ 0 \,. \label{chgconstr1}
\end{equation}
We will not analyze this further here, but we will examine the solution space in more generality  in the next section and we will show that there are very non-trivial constraints on the solution space for the non-BPS system.

\section{The solution spaces in terms of charges}
\label{SolSpace}
 
\subsection{The quartic constraint}
\label{quarticgen}

To find restrictions on the solution space and charges it turns out to be easier to reverse the usual perspective and treat the bubble equations  (\ref{simpbubble1})--(\ref{simpbubble3}) as constraints that fix $\alpha$, $\beta$ and $\gamma$ for given locations, $a_j$, of the supertubes\footnote{It is partially for this reason that we chose to reduce the configuration to three independent charge parameters.}.  

To do this most efficiently, one solves  equations (\ref{simpbubble1}) and (\ref{simpbubble3})  for $\alpha$ and $\gamma$.  Substituting this into  (\ref{simpbubble2}) generates quartic in $\beta$ that is required to vanish.  This quartic is extremely complicated and details are given in Appendix \ref{appendixA}.  On the other hand, in the scaling limit the quartic simplifies dramatically.   To that end, we substitute 
\begin{equation}
 a_1    ~=~ a_3 ~+~ \lambda \, y_1  \,,  \qquad  a_2   ~=~ a_3 ~+~ \lambda \, y_2  \label{expansion4}
\end{equation}
and expand the quartic to leading order in small $\lambda$.  We find that the result starts at $\cO(\lambda^4)$ and is actually a quadratic in  $\beta^2$:  The $\beta$ and $\beta^3$ terms actually vanish as $\cO(\lambda^6)$.   Dropping some overall factors, this quadratic, at leading order in $\lambda$, is:
\begin{eqnarray}
P_4    &=& (q+ h a_3)^2 \, y_2^2  \, \beta^4  ~+~16\,a_3^2 \, d^4 \, (Y+4)^2 \, (y_1 -y_2)^2 \nonumber \\
&& ~+~4 \,a_3  \, d^2  \, (q+ h a_3) \, \big( (y_1^2 +2\, y_1 y_2 -2 y_2^2) Y  - 4 \, (y_1^2 - 2 \,y_1 y_2 + 2\, y_2^2)  \big)\, \beta^2 \label{quadratic1}
\end{eqnarray}
The discriminant of this quadratic is:
\begin{equation}
\Delta   ~=~  16\,a_3^2 \, d^4 \,  (q+ h a_3)^2 \,y_1^2 \, (4-Y)\, \big(4\, (y_1 -2\,y_2)^2  -  (y_1^2 +4 \,y_1 y_2 -4 \,y_2^2)\,  Y  \big) \,.  \label{discriminant1}
\end{equation}

First observe that for $Y=0$ one has 
\begin{equation}
\Delta   ~=~  256\,a_3^2 \, d^4 \,  (q+ h a_3)^2 \,y_1^2 \, (y_1 -2\,y_2)^2 \,,  \label{discriminant2}
\end{equation}
 which is a perfect square and hence the quadratic always has roots.  Indeed, the quadratic has roots:
\begin{equation}
\beta^2   ~=~  \frac{16\,a_3  \, d^2 }{ (q+ h a_3)} \,, \qquad  \beta^2   ~=~  \frac{16\,a_3  \, d^2 }{ (q+ h a_3)\, y_2^2}\, (y_1 -y_2)^2 \,.   \label{BPSroots}
\end{equation}

For the non-BPS system one has, from (\ref{Ytildedefn}):
\begin{equation}
Y   ~=~  \widetilde Y ~=~   \frac{4\,h \, q \, d^2 }{ (q+ h a_3)^3} ~+~ \cO(\lambda) \,.   \label{Ylimit}
\end{equation}
Define 
\begin{equation}
 \mu ~\equiv~   \frac{y_2 }{ y_1} \,,  \qquad \qquad   \Lambda ~\equiv~   \frac{ h \, q \, d^2 }{ (q+ h a_3)^3} \,,  \qquad \qquad      f(\mu)  ~\equiv~    \frac{(1-2\,\mu)^2 }{(1+ 4\,\mu- 4\,\mu^2)}  \,,  \label{Lambdamu}
\end{equation}
then the discriminant is non-negative if either 
\begin{equation}
\Lambda  ~\le~  1 \,,  \qquad \qquad     f(\mu) ~\ge~  \Lambda   \,,   \label{range1}
\end{equation}
or
\begin{equation}
\Lambda  ~\ge~  1 \,,  \qquad \qquad   f(\mu)  ~\le~  \Lambda   \,.   \label{range2}
\end{equation}

However, since this is a quadratic in $\beta^2$, one must also show that there are positive roots.  Since the coefficient of $\beta^4$ and $\beta^0$ are non-negative,  it follows that the roots of the polynomial are either both positive or both negative.   The sign of these roots is determined by the sign of the coefficient of $\beta^2$.   This coefficient may be written:
\begin{equation}
-   2 \,a_3  \, d^2  \, (q+ h a_3) \, \big(y_1^2\, (4-Y) ~+~ \big(  4 \, (y_1- 2 \, y_2)^2 -(y_1^2 +4\, y_1 y_2 -4\, y_2^2) Y  \big)  \big) \,,  \label{coeffbetasq}
\end{equation}
which is negative in the region defined by  (\ref{range1}) and positive for (\ref{range2}).  This means that the solutions for $\beta^2$ are {\it both} positive and hence there are four real roots in the region defined by (\ref{range1}) while   (\ref{range2}) corresponds to four purely imaginary roots for $\beta$ and is, therefore, unphysical.  Since we want all the asymptotic charges to be positive, (\ref{Zasympinfty}), will generically mean we want to focus on the solutions with $ \beta >0$.  

Thus we conclude that the physical branches are defined by  (\ref{range1}) and there are always two positive, real solutions  for $\beta$ in this range.  Indeed, these solutions must correspond to the non-BPS extensions of the geometric and arithmetic branches.  We therefore focus on this domain.

\begin{figure}[t]
 \centering
    \includegraphics[width=6cm]{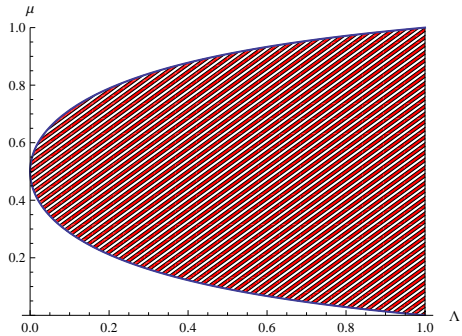}
    \caption{\it \small This shows the possible locations of the non-BPS supertubes, as determined by $\mu \equiv y_2/y_1$, as a function of the supersymmetry-breaking parameter, $\Lambda$.  The shaded area depicts the forbidden region.}
\label{fig0a}
\end{figure}

In order to preserve the proper ordering of the points, (\ref{reqinequal}), one must have $0 < \mu < 1$ and in this range one has $0 <  f(\mu) < 1$ however there are then ``forbidden regions'' for $\mu$ in which (\ref{range1}) is not satisfied.  These regions are defined by:
\begin{equation}
\frac{1}{2} - \sqrt{\frac{\Lambda}{2\,(1+\Lambda)}} \ \  ~<~   \ \ \mu \ \  ~<~  \ \  \frac{1}{2} +  \sqrt{\frac{\Lambda}{2\,(1+\Lambda)}}       \label{forbidden}
\end{equation}
and they are depicted in Fig. \ref{fig0a}.   Inside these regions the values of $\beta$ are all  complex and no real, physical solution exists.

\goodbreak

The simplest way to describe the moduli space of physical, non-BPS solutions is to start at $q=0  \Rightarrow \Lambda =0$ and slowly turn on  the supersymmetry breaking parameter, $q$.  From this we learn several things.  First, the critical control parameter is $d/\ell$, where $\ell$ is the curvature length scale at the location of the ring. (Flat space has $\ell = \infty$.)  For a given ring dipole charge, the first inequality in (\ref{range1}) tells us that a solution will cease to exist if the curvature gets too strong at the ring.  The second inequality then tells us that if the curvature is in the range that allows solutions  then $\mu$ must either be close to zero or it must be close to $1$ and $\mu =\frac{1}{2}$ is always excluded.  In other words, for the non-BPS solution with intermediate curvatures, the middle point ($a_2$) must be close to one or other of the outer points ($a_1$ and $a_3$) and as $\Lambda \to 1$ one must have $a_2 \to a_1$ or $a_2 \to a_3$. 

Therefore, if one imagines turning on the curvature, or supersymmetry breaking, slowly from flat space by increasing $q$ from zero, one starts from a situation where all points in the range $a_1 > a_2 >a_3$  are allowed but as the curvature increases the forbidden zone grows, driving $a_2$ to one or other end of the interval and then, when the curvature reaches a critical value, the scaling solutions cease to exist altogether.  Supersymmetry breaking thus forces two of the supertubes to come together and merge if the supersymmetry breaking becomes strong enough. For our choice of scaling conditions (\ref{scalingcond}), to keep the proper order of charges we need $a_2 \to a_1$ so that (\ref{geomadditional}) is satisfied. Once again for the case $a_2 \to a_3$, the order of charges has to flip and so we also need to flip the signs in the scaling conditions (\ref{scalingcond}).

\subsection{Some numerical examples}
\label{NumEx}
 
To illustrate the foregoing analysis, we fix the supertube dipole moments and positions and fix the geometry at infinity and then examine the space of solutions for the charges $\alpha, \beta$ and $\gamma$ as we vary $q$. Note that  for $q=0$ and for $q \to \infty$ the geometry is  flat and hence supersymmetric.  In both of these limits we expect the BPS and non-BPS solutions to coincide.

\begin{figure}[t]
 \centering
    \includegraphics[width=6cm]{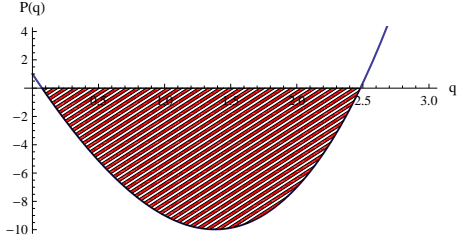}
    \caption{\it \small The cubic inequality (\ref{cubic1}) forbids non-BPS solutions for a range of $q$ determined by the shaded region of this graph.  }
\label{fig0b}
\end{figure}

We choose the following values of parameters
\begin{equation}
d=1\,, \quad  h=1\,, \quad  a_3=1\,, \quad |a_2-a_3|=2 \times10^{-8}\,, \quad|a_1-a_2| =1 \times 10^{-8}\,.  
\label{Nlayout}
\end{equation}
One should note this corresponds to having $\mu = 2/3$ in (\ref{Lambdamu})  and that this lies outside the forbidden region for small $q$.  Indeed, this configuration lies in the forbidden region for $\Lambda >  {1 \over 17}$ and with the choices (\ref{Nlayout}), this corresponds to
\begin{equation}
P(q) ~\equiv~ (q+1)^3 ~-~ 17\, q~<~  0 \,. 
\label{cubic1}
\end{equation}
This cubic inequality is depicted in Fig. \ref{fig0b} and it forbids a range of $q$ that is approximately given by $0.07258 < q <2.4808$ \footnote{The NUT charge, $q$, is, of course, quantized and is required to be a non-negative integer and so the intermediate real values of $q$ are not  physical. On the other hand, the detailed restrictions  on $q$  are a direct consequence of choosing $h=a_3 =1$ and different choices of these parameters will scale the allowed ranges of $q$ and these ranges can certainly be arranged to contain physically sensible, positive integer values for $q$ on both sides of the inequality.}.  The solution exists for smaller or larger values of $q$.

\goodbreak

\begin{figure}[!ht]
\begin{center}
 \subfigure[BPS Arithmetic Branch ]{\includegraphics[angle=0,
width=0.4\textwidth]{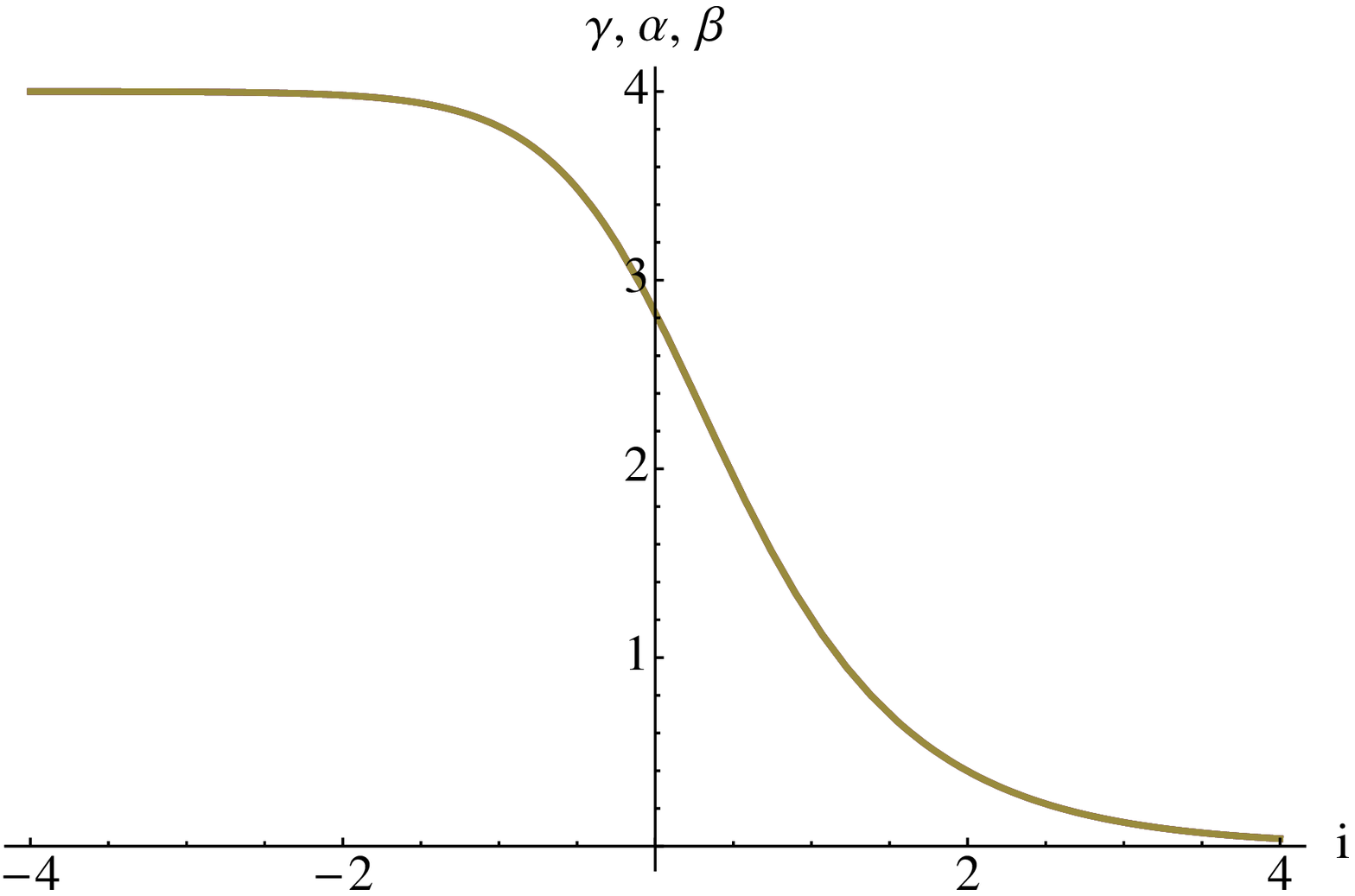} \label{fig3a}}
\subfigure[BPS Geometric Branch]{\includegraphics[angle=0,
width=0.4\textwidth]{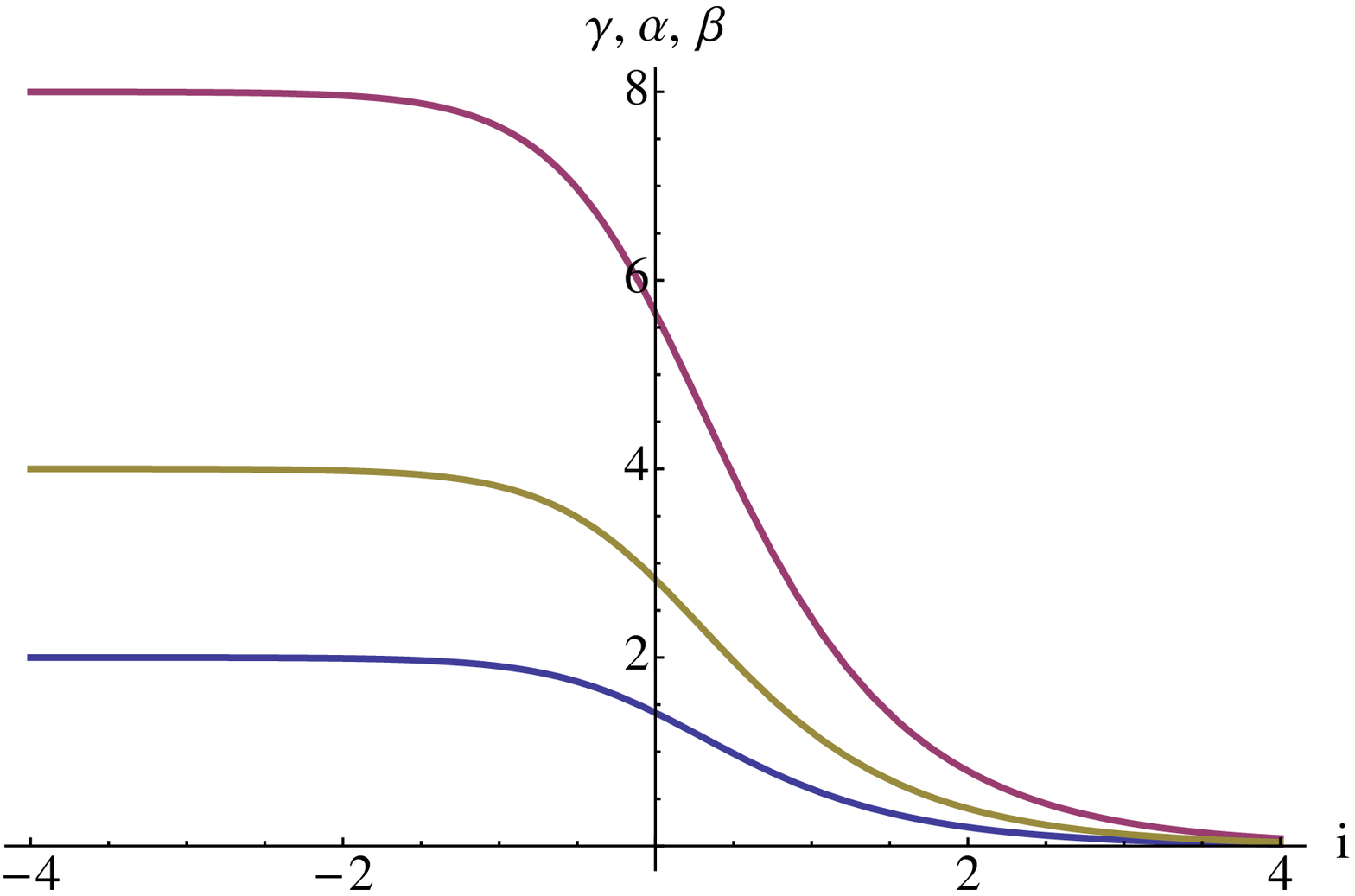}  \label{fig3b}}
\caption{{\it  \small The charges $\alpha$, $\beta$, $\gamma$ for the BPS system as a function of $i=log_{10} q$. These two graphs show the arithmetic and geometric branches of the solution.  Note  that the charges are almost identical on the arithmetic branch.}}
 \label{fig3}
\end{center}
\end{figure}

\begin{figure}[!ht]
\begin{center}
 \subfigure[non-BPS Arithmetic Branch ]{\includegraphics[angle=0,
width=0.4\textwidth]{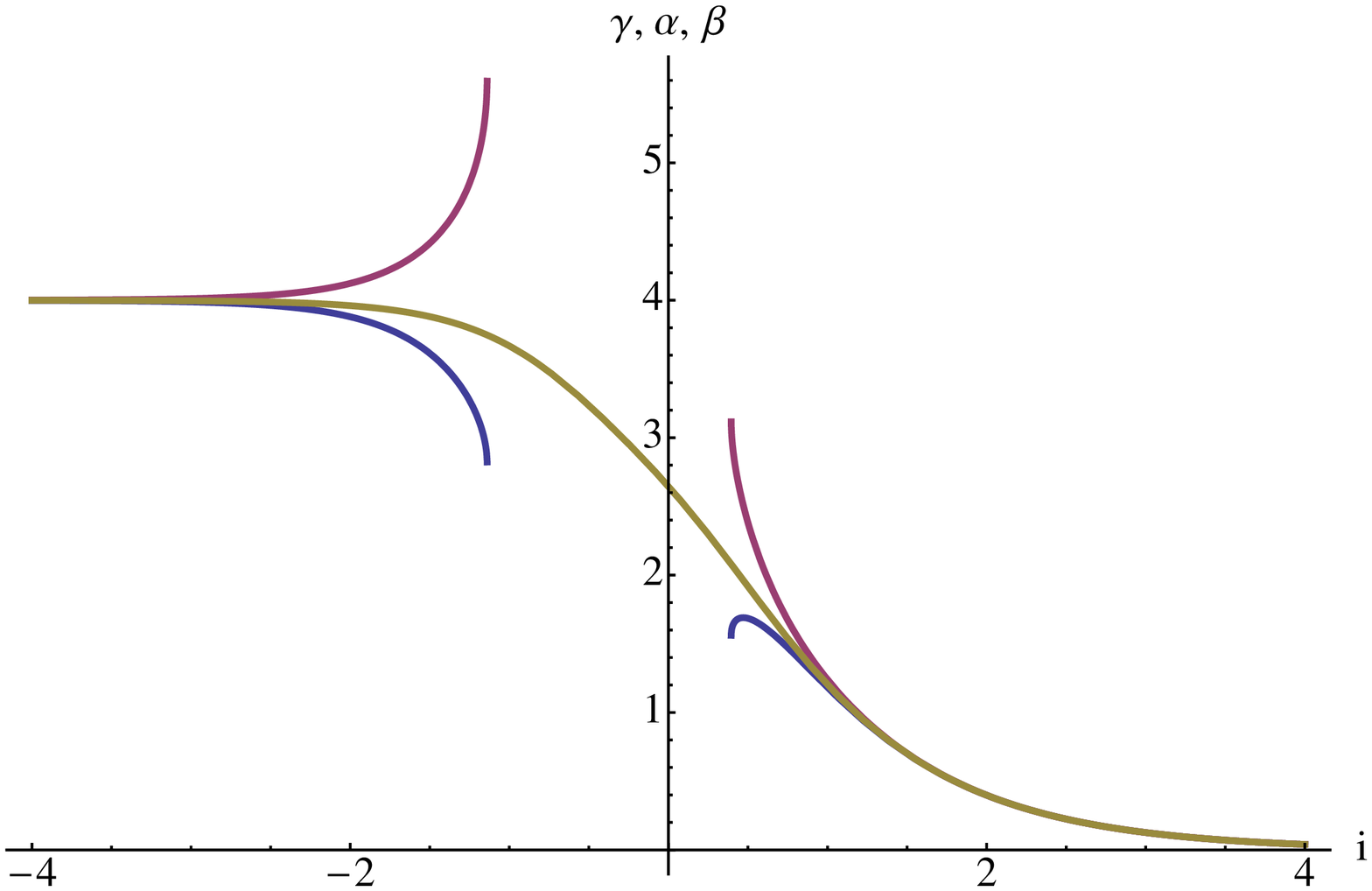} \label{fig4a}}
\subfigure[non-BPS Geometric Branch]{\includegraphics[angle=0,
width=0.4\textwidth]{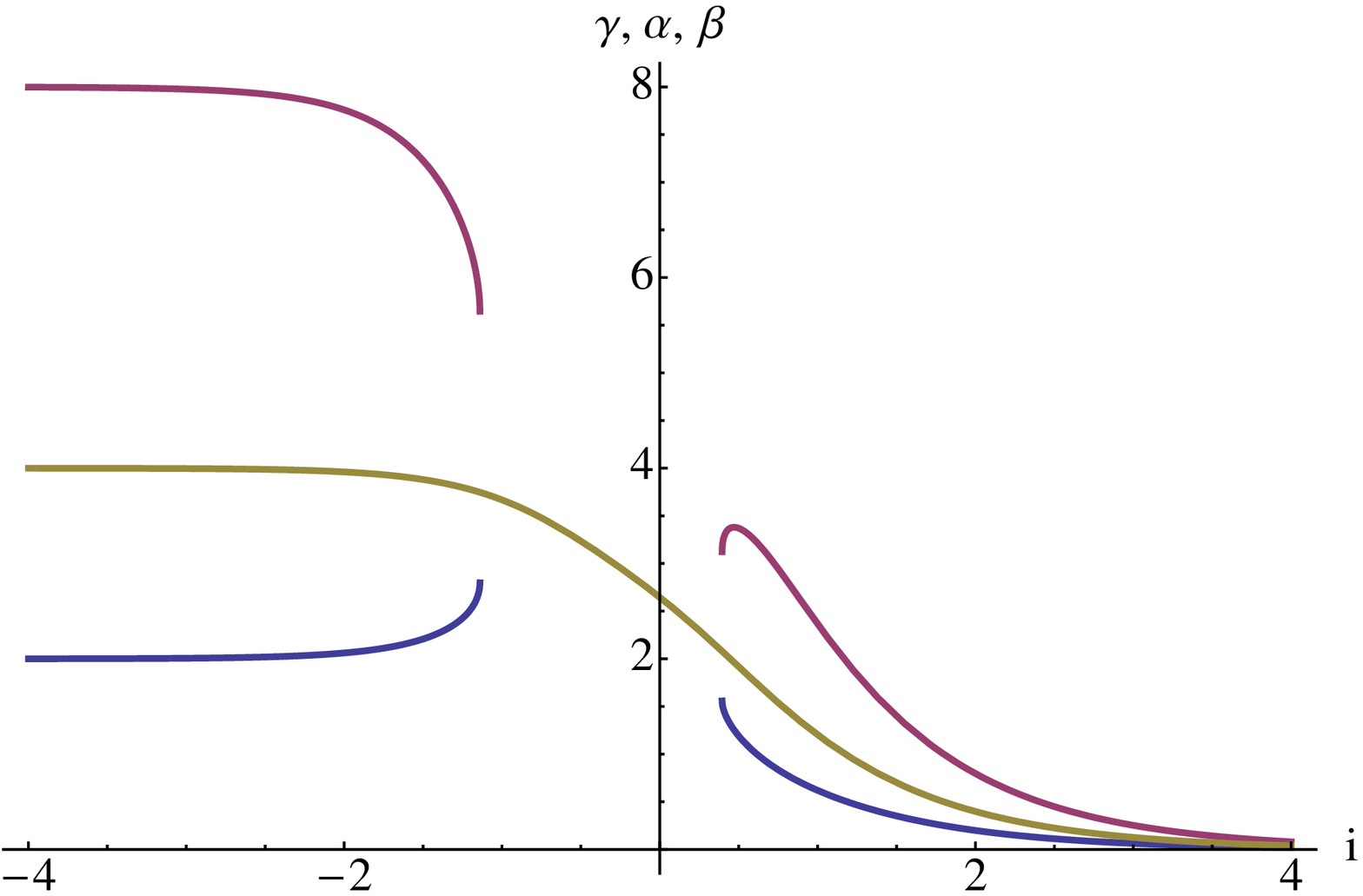}  \label{fig4b}}
\caption{{\it   \small The charges $\alpha$, $\beta$, $\gamma$  for the non-BPS system as a function of $i=log_{10} q$.  Again we show the arithmetic and geometric branches of the solution.  There is a gap in solution space  where $\beta$ and $\gamma$ become complex.  The arithmetic and geometric branches connect precisely at the gap.}}
\label{fig4}
\end{center}
\end{figure}

In Figures \ref{fig3} and \ref{fig4} we plot the solution for $\alpha, \beta$ and $\gamma$ of the exact system of the bubble equations.  We generate these solutions using the reduction of the bubble equations provided by (\ref{quartic}) and (\ref{restofcharges}).

\begin{figure}[!ht]
 \centering
    \includegraphics[width=6cm]{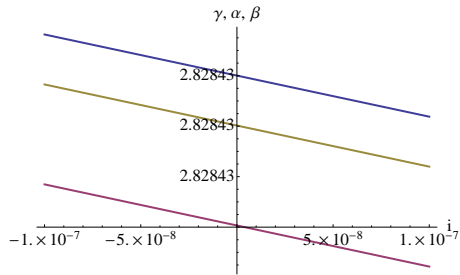}
    \caption{\it \small A close-up of Fig. \ref{fig3a} near the region $i=0$ ($q=1$). Note that the fluxes are very small but  non-zero.}
\label{fig5}
\end{figure}

The BPS and non-BPS solutions both exhibit arithmetic and geometric branches and, as one would expect, the corresponding BPS and non-BPS solutions are very similar when the NUT charge is very small or very large.  On the arithmetic branch the charges seem to be identical but a closer examination reveals that the  fluxes, $\Gamma_{ij}$, are very small, but non-zero.  Indeed, near the region $q=1$ for the BPS solution they are of the order $10^{-6}$.  We show this  detail of the BPS arithmetic branch in Fig. \ref{fig5}.   As one approaches $q \to 0$ or $q \to \infty$ the fluxes limit to exactly zero, consistent with (\ref{quarticarithmetic}).

For the non-BPS solutions we observe a ``gap'' in the solution space,  corresponding to the forbidden region identified earlier.  In this gap the values of the charges $\beta$ and $\gamma$ become complex. Somewhat surprisingly, the charge $\alpha$, which is approximately the geometric or arithmetic average of the other two charges, remains real for all values of $q$.  At leading order (\ref{arithbrnch2}) and (\ref{geombrnch2a}) then imply that $\beta$ and $\gamma$ are complex conjugates of one another but this is not true in general as can be seen at linear order from  (\ref{geombrnch3}) and at all orders from  (\ref{restofcharges}).

\begin{figure}[!ht]
\begin{center}
 \subfigure[BPS Discriminant vs. q]{\includegraphics[angle=0,
width=0.4\textwidth]{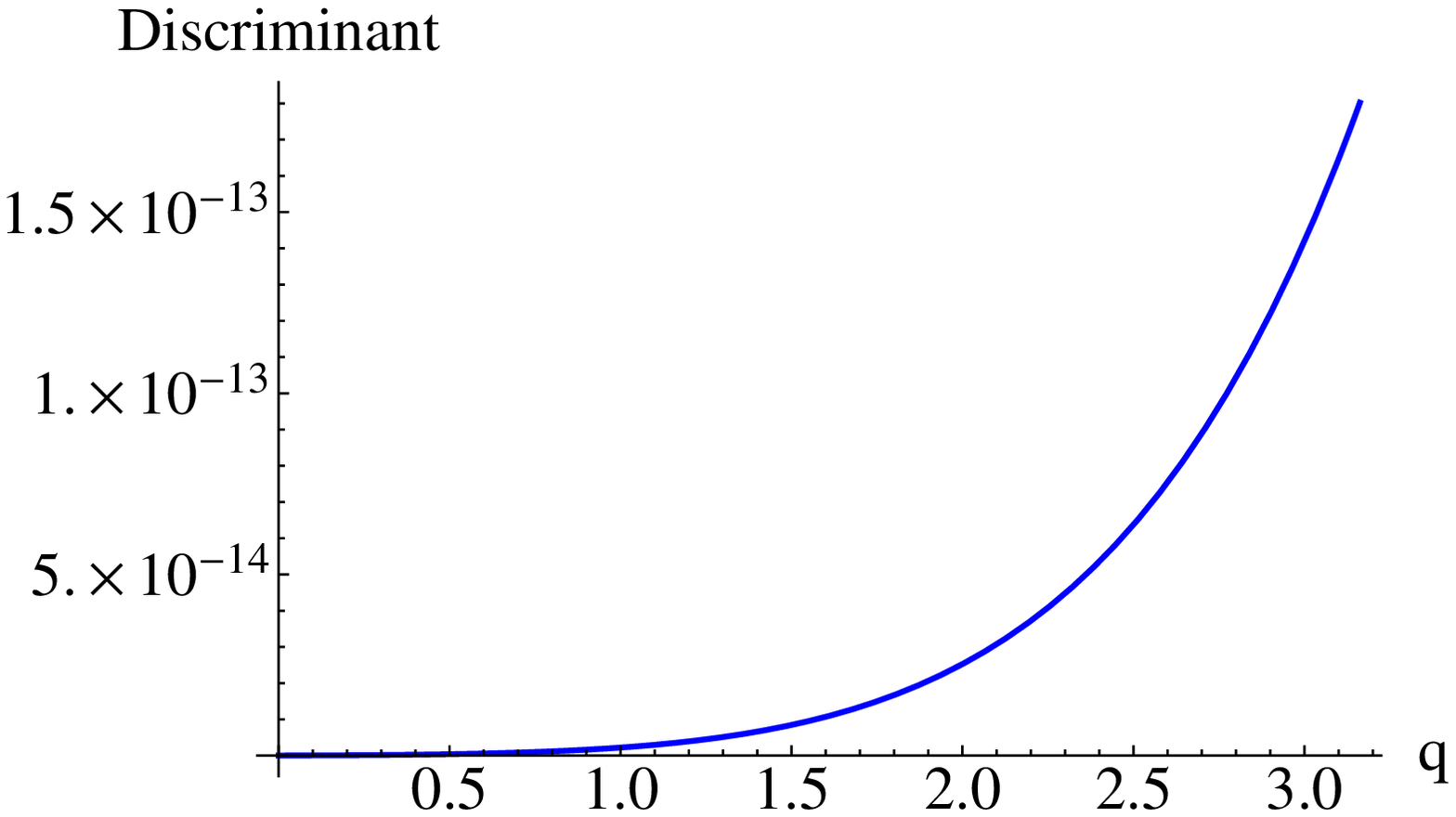} \label{fig6a}}
\subfigure[BPS Discriminant detail near $q=0$]{\includegraphics[angle=0,
width=0.4\textwidth]{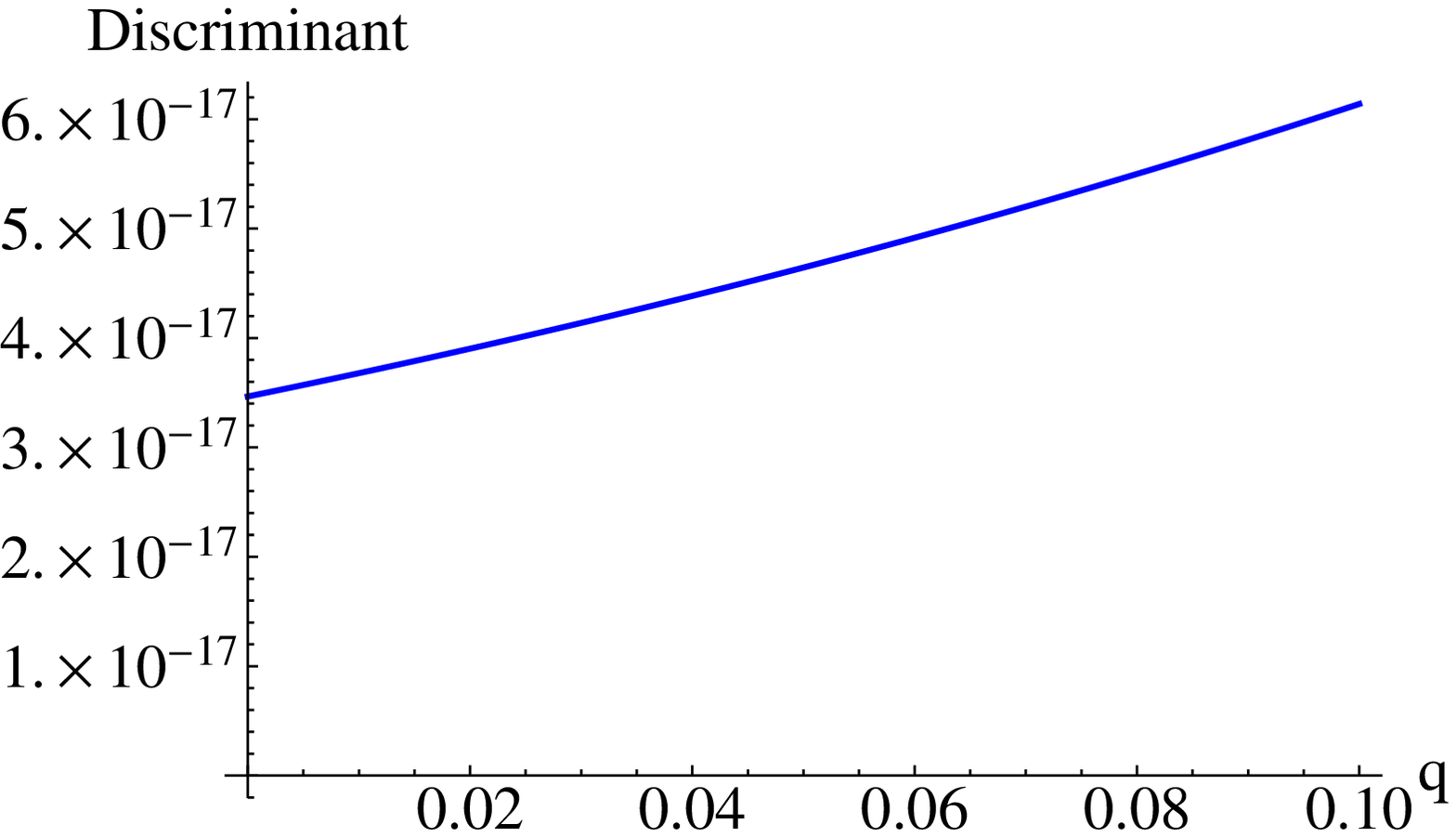}  \label{fig6b}}
\caption{{\it   \small The BPS discriminant is always positive.}}
\label{fig6}
\end{center}
\end{figure}

\begin{figure}[!ht]
\begin{center}
 \subfigure[non-BPS Discriminant vs q]{\includegraphics[angle=0,
width=0.4\textwidth]{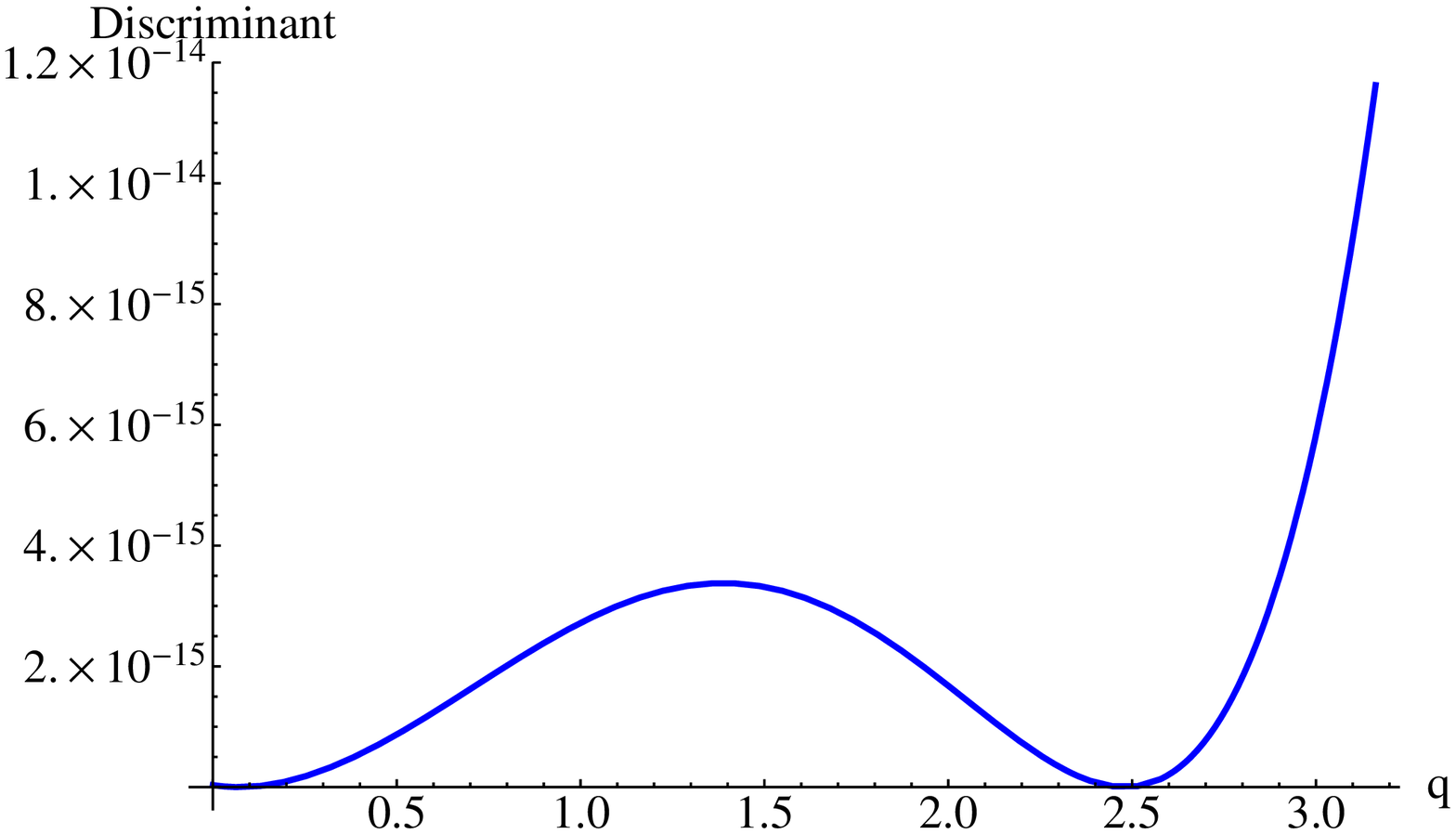} \label{fig7a}}
\subfigure[non-BPS Discriminant detail near $q=0$]{\includegraphics[angle=0,
width=0.4\textwidth]{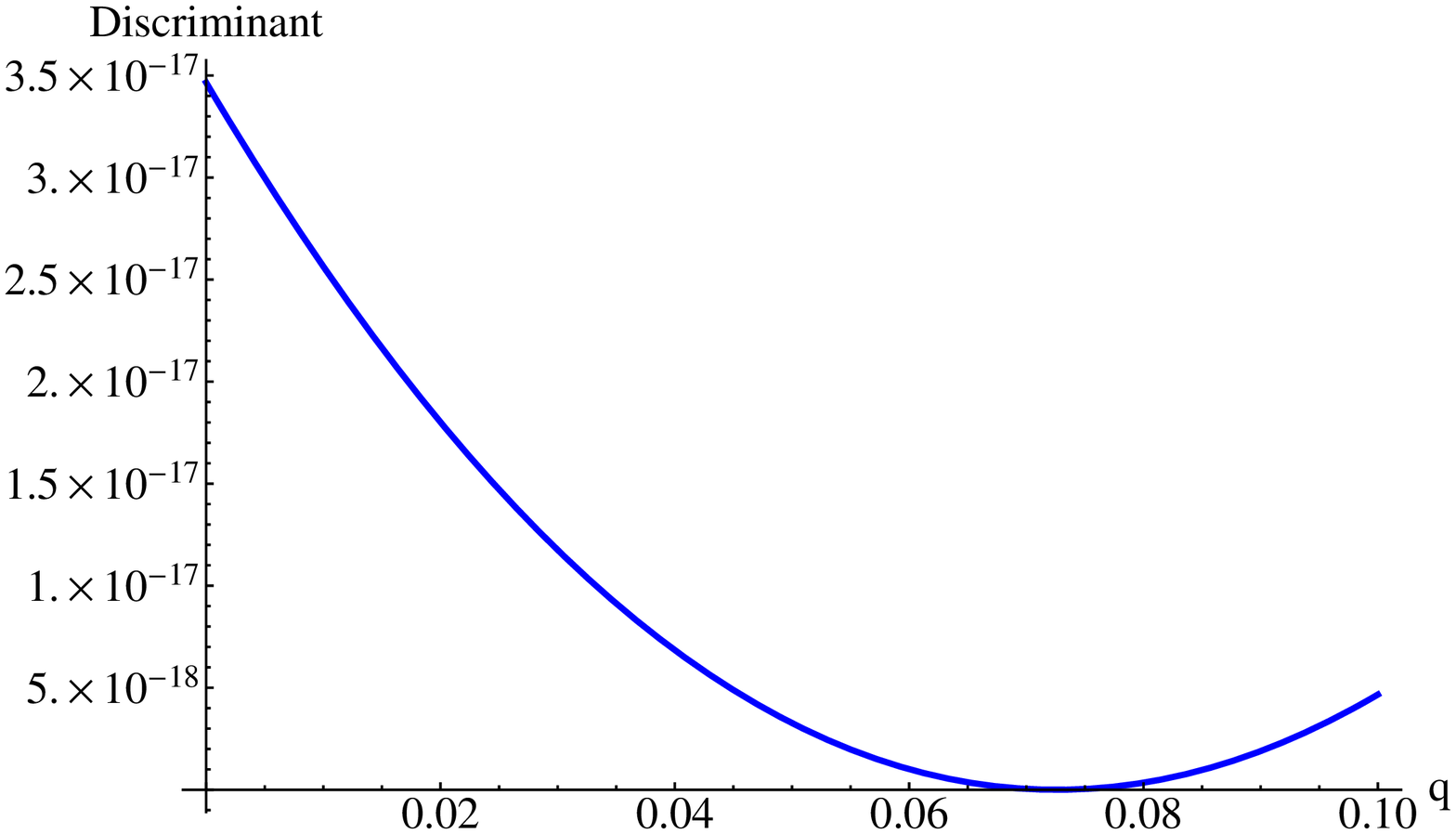}  \label{fig7b}}
\caption{{\it  \small The non-BPS discriminant is non-negative but has two  zeroes that define the edges of the forbidden region, or gap.  These zeroes closely  match the approximation given in  Fig. \ref{fig0b}.}}
\label{fig7}
\end{center}
\end{figure}

For completeness we also plot the discriminant of the full quartic in Fig. \ref{fig6} and Fig. \ref{fig7}. The discriminant of a quartic with real coefficients is positive when either all roots are real or all roots are complex. The discriminant is negative when there are two real roots and a pair of complex conjugate roots.  Fig. \ref{fig6} shows that the BPS   discriminant is always strictly positive. Since one has  real values for the charges at $q=0$, it follows that by continuously deforming $q$ one will always have real roots. The non-BPS discriminant is never negative but there are two points where it vanishes.  This is precisely where the four roots change from all  being real to all being complex. The intermediate positive regime corresponds to the region where all roots are complex. The points  where the discriminant vanishes coincide with the boundaries of the forbidden region and match the results  from (\ref{cubic1}) within the given accuracy. 

Finally we note that as $q \to \infty$ all the electric charges of the supertube go to zero.  The fact that this must happen is an immediate consequence of the radius relations,  (\ref{BPSRreln}) and (\ref{nonBPSRreln}).  We have fixed the dipole moment and the location of the supertubes and as we increase $q$ the circumference of the $U(1)$ fiber at a fixed location goes to zero.  The only way such fixed supertubes can have a circumference that limits to zero is if the electric charges of the supertubes vanish as well.

\goodbreak
 %
\goodbreak

\goodbreak
 
 %
\goodbreak

 %
\goodbreak

\section{The asymptotic structure and charges}
\label{AsympStr}

The general scaling geometries discussed in Sections \ref{Sols1}  and \ref{ScalingSols} look, at larger scales, like a black-ring wrapping the  Taub-NUT fiber and, as a result, look like a black hole in the four-dimensional, non-compact space time.  Here we complete this description by examining the asymptotic structure.  For the general configuration of supertubes described in  Section \ref{Sols1}  we give the remainder of the solution and compare it with black-ring solution and we calculate the asymptotic charges.  It should be remembered that the complete solution isn't simply an isolated black hole but is a two centered configuration with a $D6$-brane at the center  of the space and so there will be contributions to the asymptotic charges from both centers and the interactions between them.

\subsection{The BPS configuration}

To complete the solution we need the components of the angular momentum vector in $\IR^3$:
\begin{equation}
\omega ~=~ - \sum_{i=1}^3 \, q \,m_i \,   \omega_{i0}~+~ \frac{1}{8}\sum_{i}\sum_{j, (a_j<a_i)}\Gamma_{ji} \, \omega_{ij}
 \label{BPSomg}
\end{equation}
with  
\begin{equation}
\label{BPSomg2}
\omega_{ij}=-\frac{r^2sin^2\theta+(r \cos\,  \theta -a_i+r_i)(r\, \cos \theta -a_j-r_j)}{(a_j-a_i)r_i r_j} \, d\phi
\end{equation}
for $a_i>a_j$.  The label $i=0$ refers to the Taub-NUT center and, in particular, $a_0=0$ , $r_0=r$. 

In the scaling limit were the distances between the supertubes are very small compared to their distance from the center of the space  we expect the configuration to look like  a black ring of radius $R$. It follows immediately from (\ref{BPSZ1})--(\ref{BPSZ3}) that the asymptotic electric charges are:
\begin{equation} 
 Q_1~=~ Q^{(1)}_{2}+Q^{(1)}_{3} \,, \qquad  Q_2~=~ Q^{(2)}_{1}+Q^{(2)}_{3} \,, \qquad   Q_3~=~ Q^{(3)}_{1}+Q^{(3)}_{2}  \,,  \label{st1brq} 
\end{equation}
The functions describing the solution of a black ring in Taub-NUT (see \cite{Bena:2005ni} for details) precisely match the ones for a system of three clustered supertubes.   In particular,  it is useful to introduce the effective angular momentum parameter:
\begin{equation} 
m_{R}~\equiv~m_1+m_2+m_3  \,,  \label{st1brj}
\end{equation}
and then note that (\ref{mjres}) and (\ref{BPSRreln}) yield the radius relation of the effective black ring:
\begin{equation} 
\label{st1RR}
m_{R}V_{R}~=~ \frac{1}{2}(k_1+k_2+k_3)
\end{equation}

In the scaling limit, to first order, we find $\omega_{10}\approx \omega_{20}\approx\omega_{30}\approx\omega_{R0}$ and $\omega_{21}\approx\omega_{32}\approx\omega_{31}$. The latter, when combined with (\ref{sumGamma}), means that the ``supertube interaction term'' in  (\ref{BPSomg})  vanishes to leading order.  This is the familiar merger condition on local angular momentum contributions \cite{Bena:2006kb}. The remaining term in   (\ref{BPSomg})  is  the sum of the angular momentum contributions for each supertube about the center of space  thus give  the correct effective $\omega$ for the black ring.  

We can compactify our solution on the Taub-NUT circle to obtain a microstate geometry for a four-dimensional  black-hole. By standard Kaluza-Klein reduction arguments and in the same spirit as in \cite{Bena:2009ev} we can write the metric as:
\begin{equation}
 \label{BR1KKmetric}
ds^2=\frac{I_4}{(Z_1Z_2Z_3)^{2/3}V^2}\Big(d\psi+A -\frac{\m V^2}{I_4}(dt+\omega)\Big)^2+\frac{V(Z_1Z_2Z_3)^{1/3}}{\sqrt{I_4}}ds_4^2 \,,
\end{equation}
where
\begin{equation}
\label{BR14dmetric}
ds_4^2=-I_4^{-1/2}(dt+\omega)^2+I_4^{1/2}(dr^2+r^2d\Omega_2^2)
\end{equation}
is the 4-dimensional Lorentzian metric and one defines: 
\begin{equation}
 \label{BR1I4}
I_4 ~\equiv~Z_1Z_2Z_3V-\m^2V^2  \,.
\end{equation}
To identify the asymptotic charges and ADM mass it is most convenient to pass to a rest-frame at infinity by  re-defining
\begin{equation}
 \label{BR1derotate}
\tilde{\psi}~=~\psi-\frac{\m_0h^2}{I_0}t  \,,
\end{equation}
where $I_0=h-h^2\m_0^2$ and $\m_0$ are the constant values of $I_4$ and $\m$ at infinity. 
For the BPS solution  one has $\m_0=m_\infty=-\frac{m_R}{R}=\frac{J_R}{16R}$.\\
 
The asympotic charges can be read from the expansion of the metric and the warp factors $Z_I$. After simplifying the resulting expressions and by making use of the radius relation. 
Thus for the electric charges we find: 
\begin{equation}
 \label{BR1Q}
\bar{Q}_I~=~Q_I \,, 
\end{equation}
and the ADM mass is given by:
\begin{equation}
 \label{BR1Mass}
M~=~\frac{1}{8I_0^{3/2}}\left(4 \, q+16m_\infty^2h^2R+h\left(Q_1+Q_2+Q_3\right)\right) \,.
\end{equation}
The Kaluza-Klein charge coming from the momentum around the $\psi$-fiber is:
\begin{equation}
 \label{BR1KKQ}
P~=~ - \frac{h^2m_\infty}{4I_0^2}\big(4hm_\infty^2 RV_R+4h  \, (R+1)+4h^2m_\infty  R+h \big(Q_1+Q_2+Q_3\big)\big)
\end{equation}
and  the angular momentum around the $\phi$ direction is:
\begin{equation}
 \label{BR1J3}
J_3~=~\frac{ qh m_\infty R}{I_0^{3/2}}( hm_\infty^2-1)~=~ -\frac{m_\infty qR}{\sqrt{I_0}} \,.
\end{equation}
The black ring horizon area is given by:
\begin{equation}
 \label{BR1horizon}
A_H~=~2\pi^2q\sqrt{J_4} \,,
\end{equation}
where $J_4$ is the $E_{7(7)}$ quartic invariant:
\begin{equation}
 \label{BR1J4}
J_4~=~ -\sum_{I=1}^3Q_I^2k_I^2+2\sum_{I<J}k_Ik_JQ_IQ_J+8k_1k_2k_3J_R \,.
\end{equation}
%
%

\subsection{The non-BPS configuration}
 
The angular momentum vector for non-BPS supertubes is still of the form (\ref{kansatz}) but now $\mu$ is given by 
(\ref{munonBPS})  and  (\ref{munonBPSdefns}).   The expression for $\omega$ can be obtained from the general expressions in  
  \cite{Bena:2009en} and for the solution in this paper we have:
\begin{eqnarray}
\label{st2omg}
\omega &=& \sum_I k_I\, \omega_I^{(1)} ~+~ h\, \sum_I\sum_{j\neq I}\frac{Q^{(I)}_{j}k_I}{4}\, \omega_{Ij}^{(3)}~+~  q\, \sum_I\sum_{j\neq I}\frac{Q^{(I)}_{j}k_I}{4}\, \omega_{Ij}^{(5)}  \\  
&& \qquad ~+~  k_1k_2k_3\, \big(h^2\omega^{(6)}+q^2\omega^{(7)}+qh\omega^{(8)} \big)~+~  \omega^{(9)} \,,
\end{eqnarray}
where
\begin{eqnarray}
\label{st2omgs}
\omega_I ^{(1)}&=& \frac{h}{2}\frac{r \cos\theta-a_I}{r_I}d\phi+\frac{q}{2}\frac{r-a_I \cos \theta}{a_Ir_I}\, d\phi \,, \\
\omega_{Ij}^{(3)}&=& \frac{r^2+a_Ia_j-(a_I+a_j)r \cos \theta}{2(a_I-a_j)r_Ir_j}\, d\phi \,,\\ 
\omega_{Ij}^{(5)} &=&  \frac{r\, (a_j+a_I \cos 2\theta)-(r^2+a_Ia_j) \cos\theta}{2a_I(a_j-a_I)r_I r_j}\, d\phi\,, \\ 
\omega^{(6)}&=&0 \,,  \qquad  \omega^{(7)}~=~  \frac{r^2 \sin^2\theta}{a_1a_2a_3r_1r_2r_3}\,d\phi , \\
\omega^{(8)}&=&  \frac{r^3+r(a_1a_2+a_2a_3+a_1a_3)-\left(r^2(a_1+a_2+a_3)+a_1a_2a_3\right) \cos\theta}{2a_1a_2a_3r_1r_2r_3}\,d\phi , \\ 
\omega^{(9)}&=&  \bigg(\kappa-m_0 \cos\theta-\sum_i m_i\frac{r \cos \theta-a_i}{r_i} \bigg)\, d\phi  \,.
\end{eqnarray}
The absence of Dirac strings and CTC's requires that the bubble equations (\ref{nonBPSbubble1})--(\ref{nonBPSbubble3}) are satisfied and that:
\begin{eqnarray} 
\kappa &=&  -q\sum_i\frac{k_i}{2a_i}~-~ h\,q\frac{k_1k_2k_3}{2a_1a_2a_3}~-~ h\sum_I\sum_{j\neq i}\frac{Q^{(I)}_{j}k_I}{8(a_I-a_j)} \label{st2ds1} \,, \\
 m_0 &=& -q\sum_i\frac{k_i}{2a_i}~-~hq\frac{k_1k_2k_3}{2a_1a_2a_3}~+~q\sum_I\sum_{j\neq i}\frac{Q^{(I)}_{j}k_I}{8a_I(a_I-a_j)}   \label{st2ds2} \,.
\end{eqnarray}

In the scaling limit, where the distances between the supertubes are very small compared to their distance from the center of the space, the solution should effectively behave as a black ring of radius $R$. Once more it is easy to see that the functions describing the solution match those of the black ring if one identifies:
\begin{equation}
 \label{st2br1}
Q^{(I)}=\sum_{i\neq I}Q^{(I)}_{i}  \,, \qquad  
m_{R}=m_1+m_2+m_3 \,.
\end{equation}
Denoting   the terms corresponding to the supertube and the black ring solution with the subscripts ``$st$'' and ``$br$'', 
we then have:
\begin{equation}
 \label{st2br3}
\ka_{br}=\tilde{\ka}_{st}=\ka_{st}+h\sum_I\sum_{j\neq i}\frac{Q^{(I)}_{j}k_I}{8(a_I-a_j)}
\end{equation}
\begin{equation}
 \label{st2br4}
m_{0,br}=\tilde{m}_{0,st}=m_{0,st}-q\sum_I\sum_{j\neq i}\frac{Q^{(I)}_{j}k_I}{8a_I(a_I-a_j)}
\end{equation}
\begin{equation}
 \label{st2br5}
\mu^{(1)}_{br}=\mu^{(1)}_{I,st} \,, \qquad  \omega^{(1)}_{br}=\omega^{(1)}_{I,st}
\end{equation}
\begin{equation}
 \label{st2br6}
\mu^{(2)}_{br}=\mu^{(3)}_{Ij,st} \,, \qquad \omega^{(2)}_{br}=\tilde{\omega}^{(3)}_{Ij,st}=\omega^{(3)}_{Ij,st}-\frac{1}{2(a_I-a_j)}d\phi
\end{equation}
\begin{eqnarray}
 \label{st2br7}
&&\mu^{(3)}_{br}=\mu^{(2)}_{st} \,, \qquad \omega^{(3)}_{br}=\omega^{(2)}_{st}\,, \qquad 
\mu^{(4)}_{br}=\tilde{\mu}^{(5)}_{Ij,st}=\mu^{(5)}_{Ij,st}-\frac{1}{2(a_j-a_I)a_Ir} , \\
&& \omega^{(4)}_{br}=\tilde{\omega}^{(5)}_{Ij,st}=\omega^{(5)}_{Ij,st}-\frac{\cos\theta}{2(a_j-a_I)a_I} \,, \qquad \mu^{(5)}_{br}=\mu^{(4)}_{st}  \,, \qquad \omega^{(5)}_{br}=\omega^{(4)}_{st} \,, \\
&&\mu^{(i)}_{br}=\mu^{(i)}_{st} \,, \qquad \omega^{(i)}_{br}=\omega^{(i)}_{st}\,, \quad  i= 6,7,8,9\,.
\end{eqnarray}

 The modified terms $\tilde{\kappa}_{st}$, $\tilde{m}_{0,st}$, $\tilde{\omega}^{(3)}_{Ij,st}$, $\tilde{\mu}^{(5)}_{Ij,st}$, $\tilde{\omega}^{(5)}_{Ij,st}$ still describe exactly the same three-supertube system because the modifications involve the re-shuffling of harmonic functions into the $\mu^{(i)} V $ with compensating shifts in $M$ (along with corresponding changes in $\omega$'s).  
 
The non-BPS  radius relation for the black ring is:
\begin{equation}
 \label{st2br11}
\hat{m}_{R}V_{R}=16\sum_I{\hat{k}_I}+16hq\frac{\hat{k}_1\hat{k}_2\hat{k}_3}{V_{R}R^3} \,,
\end{equation}
which is to be identified with (\ref{nonBPSRreln}), 
where 
\begin{equation}
\label{effectivem}
\hat{m}_i=m_i \left(h+\frac{q}{|a_i|}\right)^{-1}
\end{equation}
is the effective angular-momentum parameter.
%

The electric charges of the non-BPS configuration are
\begin{equation}
 \label{br2Q}
\bar{Q}_I~=~ Q_I+\frac{4q}{R^2}\frac{C^{IJK}}{2}k_Jk_K  \,,
\end{equation}
while the ADM mass is given by
\begin{equation}
 \label{br2ADM}
M~=~\frac{1}{I_0^{3/2}}\left(\frac{\, q}{2}+\frac{h}{8}\left(\bar{Q}_1+\bar{Q}_2+\bar{Q}_3\right)-m h\,  \left(k_1+k_2+k_3\right)  \right) \,.
\end{equation}
The Kaluza Klein charge is now
\begin{equation}
 \label{br2KKQ}
P=\frac{1}{I_0^2} \Big( h^2(h +m^2)\,  \left(k_1+k_2+k_3\right) - \coeff{1}{4}\, mh^2\,\left(\bar{Q}_1+\bar{Q}_2+\bar{Q}_3\right)-m^3q \Big)
\end{equation}
and the angular momentum is:
\begin{equation}
 \label{br2J3}
J_3=\frac{q}{2\sqrt{I_0}} \Big(\left(k_1+k_2+k_3\right)+\frac{1}{R}\,  \vec k \cdot \vec Q +\frac{1}{R^2} \big(h+\frac{2q}{R} \big)k_1k_2k_3 \Big) \,.
\end{equation}

For the horizon area we find:
\begin{equation}
 \label{br2horizon}
\hat{A}_H=2\pi^2q\sqrt{\hat{J}_4} \,,
\end{equation}
where 
\begin{equation}
 \label{br2J4}
\hat{J}_4=-\sum_{I=1}^3Q_I^2\hat{k}_I^2+2\sum_{I<J}\hat{k}_I\hat{k}_JQ_IQ_J-8\hat{k}_1\hat{k}_2\hat{k}_316\hat{m}_R \,,
\end{equation}
which, upon setting $16\hat{m}_R=-J_R$, matches the horizon area of a BPS black ring with dipole and angular momentum parameters equal to the effective ones.

\section{Conclusions}
\label{Conclusions}

The non-BPS solutions obtained using the ``floating brane'' Ansatz have greatly enriched the families of known extremal solutions.  The fact that these solutions are also determined by solving linear systems of equations has also enabled one to generate families of multi-centered solutions.  To date, the majority of such solutions have involved black holes or multi-centered black rings, however, it is evident from the work presented here that one can use this approach to generate interesting families of non-BPS scaling microstate geometries.    By using multiple species of supertubes one can generate scaling geometries whose appearance at large scales exactly matches that of a black hole or black ring and yet, via spectral flow, these solutions represent true microstate geometries in that they are smooth, horizonless and have a long $AdS$ throat.

As we have shown here,  one of the interesting new features of these non-BPS microstate geometries is that the breaking of supersymmetry places restrictions on the moduli space of solutions, compared to the analogous BPS solutions.  If the supersymmetry breaking scale is large in the vicinity of the supertubes then a smooth solution does not exist  and when the supersymmetry breaking is sufficiently small then there are still bounds upon the allowed moduli.  

The restrictions on parameters that are imposed by supersymmetry breaking emerge in a very interesting manner: The equations of motion can be solved for all values of the parameters and the restrictions then emerge from the bubble equations, which require that there are no closed timelike curves near the supertube.   While it has not been shown in general, there is good evidence that the bubble equations can be thought of as a condition of finding a local minimum of a potential for the supertube interaction.  That is, the interactions between supertubes involve non-trivial forces between their magnetic dipoles and electric charges and the requirement of a static solution forces them to find an equilibrium configuration.   More generally, a dynamic solution will involve some kind of oscillation around this equilibrium configuration\cite{Marolf:2005cx, Bena:2005zy}.   One way one can interpret the results presented here is that the supersymmetry-breaking holonomy modifies these forces and the potential and restricts the range of equilibrium configurations and indeed wipes out the minima if the supersymmetry breaking is too strong.

There is also a rather natural aspect to the restriction placed upon the bubbles, discussed in Section \ref{quarticgen},   that shows that supersymmetry breaking forces two of the supertubes to come together.   We have described the supersymmetry breaking in terms of the supertube supersymmetry being broken by holonomy, but the supersymmetry breaking is a rather more democratic between the elements of the solution.  One should recall that each supertube is a ${1 \over 2}$-BPS state and that the Taub-NUT background is similarly a  ${1 \over 2}$-BPS state.  Any three out of the four of these elements will preserve four supersymmetries and create a ${1 \over 8}$-BPS state but the whole point of the non-BPS procedure is that all four elements, when put together, do not agree on which supersymmetries to preserve and thus the supersymmetry is thus completely broken and the control parameters are the separation of the different geometric elements.  Thus the solution is essentially supersymmetric when there are only three elements involved in determining the background.  Given a strong background holonomy, one can preserve some approximate supersymmetry locally by bringing two supertubes together while maintaining a much larger, finite separation from the third. Thus one can think of the restriction on the moduli space in terms of the solution trying to preserve approximate supersymmetry locally by forcing supertubes into clusters and excluding regions of moduli space in which all the supertubes are widely separated with large intermediate regions in which the supersymmetry is broken.  Whether this perspective is born out in more general classes of solution remains to be seen but it is certainly worthy of further investigation.

More generally, even though we have studied an extremely simple example, it is very tempting to conjecture other implications for more general classes of non-BPS microstate geometries.   It may be that the natural geometric elements of microstate geometries are ${1 \over 2}$-BPS ``atoms''  that do not necessarily preserve any supersymmetry when combined together.     The moduli space of such non-BPS solutions would then be restricted so that in each region some approximate supersymmetry would survive. This would necessitate that the ``atoms'' have strong enough local charges  (like the dipole moment, $d$, in $\Lambda$ in (\ref{Lambdamu})) to open up a non-trivial moduli spaces in the presence of the other charges. This would also be balanced against the separation between the geometric elements.   If one were to think of microstate geometries being nucleated in regions of high curvature then this would favor the initial formation of bubbles with larger dipole charges.  As this happened, the nucleation of bubbles would  involve diluting the curvature over a larger region thus enabling bubbles with smaller dipole moments to occur through either splitting of bubbles or nucleation.  (In a Taub-NUT space, this curvature dilution would  involve a transition to multi-centered Taub-NUT so that the background curvature would be distributed into a cloud around all the centers.)   This is consistent with the belief, common to both the fuzzball proposal and ideas of emergent space-time, that the microstate structure of a black hole is not localized in a Planck-scale region around a classical singularity, but is, instead, smeared out over a region whose scale is set by the horizon area.

While the solutions we have considered here are far too simple to provide one with the general picture of non-BPS microstate geometries, we have shown that there are interesting families of non-BPS, scaling microstate geometries and we have shown how supersymmetry breaking can modify and restrict those families in very interesting ways that might naturally be characterized in terms of some locally approximate supersymmetry.

\bigskip
\bigskip
\bigskip
\leftline{\bf Acknowledgements}
\smallskip
This work was supported in part by DOE grant DE-FG03-84ER-40168.    We would like to thank Iosif Bena for valuable discussions. OV would like to thank the USC Dana and David Dornsife College of Letters, Arts and Sciences for support through a College Doctoral Fellowship.




\appendix 
\section{Details of the quartic}
\label{appendixA}
\renewcommand{\theequation}{A.\arabic{equation}}
\renewcommand{\thetable}{A.\arabic{table}}
\setcounter{equation}{0}

The bubble equations can be written in a more symbolic form as:
\begin{eqnarray}
 \label{regfull3}
g(\bb-\al)+b(\g-\bb)+d_1\bb\g+f_1&=& 0  \,, \\ 
c(\al-\g)+g(\al-\bb)+d_2\al^2+f_2&=&  0  \,, \\ 
b(\bb-\g)+c(\g-\al)+d_3\bb\g+f_3&=&  0 \,, 
\end{eqnarray}
with the following  definitions:
\begin{eqnarray}
&& g=\frac{1}{|a_1-a_2|}\,,   \qquad b=\frac{1}{|a_1-a_3|}\,,  \qquad c=\frac{1}{|a_2-a_3|}\,,  \qquad d_i=\frac{V_{a_i}}{4d^2}  \label{quarticident1} \\
&& V_{a_i}=\left(h+\frac{q}{a_i}\right)\,,   \qquad f_1=f_3=-4-Y\,, \qquad f_2=-4+Y \,. \label{quarticident2}
\end{eqnarray}
Summing the bubble equations  we get the relatively simple condition:
\begin{equation}
 \label{regRRfull3}
\bb\g(d_1+d_3)+ d_2\al^2+f=0 \,,
\end{equation}
where $f=f_1+f_2+f_3$. For later convenience we define the following:
\begin{eqnarray}
 \label{quarticsymbols} 
A&=&  c\left(1-\frac{b}{g}\right)-b \,, \qquad B= \frac{cd_1}{g}-d_3  \,, \qquad  \Gamma = \frac{cf_1}{g}-f_3 \,,  \\
\Psi&=& 1-\frac{b}{g} \,, \qquad  \Phi= \frac{2\, b}{g}- \frac{2\,  b^2}{g^2}+ \frac{2\, d_1 f_1}{g^2}+\frac{d_1}{d_2}+\frac{d_3}{d_2}  \,.
\end{eqnarray}
The foregoing system of equations can be explicitly solved by reducing it to a quartic polynomial  for the charge  $\beta$:
\begin{equation}
 \label{quartic}
s_4\bb^4+s_3\bb^3+s_2\bb^2+s_1\bb+s_0~=~ 0 \,,
\end{equation}
from which $\alpha$ and $\gamma$ can be determined using:
\begin{equation}
 \label{restofcharges}
\gamma ~=~ \frac{\G+A\bb}{A-B\bb} \,, \qquad  \alpha ~=~ \beta + \frac{b}{g}(\g-\beta)+\frac{d_1}{g}\bb\g+\frac{f_1}{g}=0 \,.
\end{equation}
The coefficients of the quartic (\ref{quartic}) are given by:
\begin{eqnarray}
s_0&=&   \left(\G\frac{b}{g}+A\frac{f_1}{g}\right)^2+A^2\frac{f}{d_2}   \,, \qquad s_4=\left(B\Psi-A\frac{d_1}{g}\right)^2  \,,
 \label{quartica1} \\
 s_1&=& 2A\G\frac{b^2}{g^2}+\G A\Phi+2\G^2\frac{bd_1}{g^2}+2A^2\frac{f_1}{g}\Psi+2\left(A^2-\G B\right)\frac{f_1b}{g^2}-2AB\left(\frac{f_1^2}{g^2}+\frac{f}{d_2}\right) \,,\label{quartica2} \\
 s_2&=& A^2\Psi^2+A^2\frac{b^2}{g^2}+\G^2\frac{d_1^2}{g^2}+\left(A^2-\G B\right)\Phi+2\G A\frac{d_1}{g}\Psi+4\G A\frac{bd_1}{g^2}-4AB\frac{f_1}{g}\Psi \nonumber \\
&& \qquad  \qquad \qquad \qquad-2AB\frac{f_1b}{g^2}+B^2\left(\frac{f_1^2}{g^2}+\frac{f}{d_2}\right) \,,  \label{quartica3}   \\
s_3&=& -2AB\Psi^2+2\G A\frac{d_1^2}{g^2}-AB\Phi+2\frac{d_1}{g}\left(A^2-B\G\right)\Psi+2A^2\frac{d_1b}{g^2}+2B^2\frac{f_1}{g}\Psi  \,.
 \label{quartica4} 
\end{eqnarray}
For  $q=0$  one has  $s_3=s_1=0$ and the quartic,  (\ref{quartic}), collapses to a quadratic in $\beta^2$.





\end{document}